# Roadmap on Data-Centric Materials Science

Stefan Bauer[1], Peter Benner[2], Tristan Bereau[3], Volker Blum[4], Mario Boley[5], Christian Carbogno[6], C. Richard A. Catlow[7], Gerhard Dehm[8], Sebastian Eibl[9], Ralph Ernstorfer[10], Ádám Fekete[11], Lucas Foppa[6], Peter Fratzl[12], Christoph Freysoldt[8], Baptiste Gault[8,13], Luca M. Ghiringhelli[6,14], Sajal K. Giri[15], Anton Gladyshev[16], Pawan Goyal[2], Jason Hattrick-Simpers[17], Lara Kabalan[7,18], Petr Karpov[9], Mohammad S. Khorrami[8], Christoph Koch[16], Sebastian Kokott[6,19], Thomas Kosch[20], Igor Kowalec[7], Kurt Kremer[21], Andreas Leitherer[6,22], Yue Li[8], Christian H. Liebscher[8], Andrew J. Logsdail[7], Zhongwei Lu[7], Felix Luong[5], Andreas Marek[9], Florian Merz[23], Jaber R. Mianroodi[8], Jörg Neugebauer[8], Zongrui Pei[24], Thomas A. R. Purcell[6,25], Dierk Raabe[8], Markus Rampp[9], Mariana Rossi[26], Jan-Michael Rost[27], James Saal[28], Ulf Saalmann[27], Kasturi Narasimha Sasidhar[8], Alaukik Saxena[8], Luigi Sbailò[11], Markus Scheidgen[11], Marcel Schloz[16], Daniel F. Schmidt[5], Simon Teshuva[5], Annette Trunschke[29], Ye Wei[30], Gerhard Weikum[31], R. Patrick Xian[32], Yi Yao[6], Junqi Yin[33], Meng Zhao[16], Matthias Scheffler[6,34]

[1] School of Computation, Information and Technology, Technical University of Munich & Helmholtz AI, Munich, Germany
[2] Max Planck Institute for Dynamics of Complex Technical Systems, Magdeburg, Germany
[3] Institute for Theoretical Physics, Heidelberg University, Heidelberg, Germany
[4] Thomas Lord Department of Mechanical Engineering and Materials Science, Duke University, Durham, North Carolina, USA
[5] Monash University, Department of Data Science and AI, Melbourne, Australia
[6] The NOMAD Laboratory at the FHI of the Max-Planck-Gesellschaft and IRIS-Adlershof of the Humboldt-Universität zu Berlin, Berlin, Germany
[7] Max Planck Centre on the Fundamentals of Heterogeneous Catalysis (FUNCAT), School of Chemistry, Cardiff University, Cardiff, UK
[8] Max-Planck-Institut für Eisenforschung GmbH, Düsseldorf, Germany
[9] Max Planck Computing and Data Facility, Garching, Germany
[10] Institute for Optics and Atomic Physics, Technical University of Berlin, Berlin, Germany
[11] Physics Department and IRIS-Adlershof, Humboldt-Universität zu Berlin, Berlin, Germany
[12] Max Planck Institute of Colloids and Interfaces, Potsdam, Germany
[13] Department of Materials, Imperial College, London, UK
[14] Department of Materials Science and Engineering, Friedrich-Alexander Universität, Erlangen-Nürnberg, Germany
[15] Department of Chemistry, Northwestern University, Illinois, USA
[16] Humboldt-Universität zu Berlin, Department of Physics, Berlin, Germany
[17] Department of Materials Science and Engineering, University of Toronto, Toronto, Canada
[18] STFC Hartree Centre, Daresbury Laboratory, Daresbury, Warrington, UK
[19] Molecular Simulations from First Principles e.V., Berlin, Germany
[20] Humboldt-Universität zu Berlin, Department of Computer Science, Berlin, Germany
[21] Max Planck Institute for Polymer Research, Mainz, Germany
[22] ICFO-Institut de Ciencies Fotoniques, The Barcelona Institute of Science and Technology, Castelldefels (Barcelona), Spain

[23] Lenovo HPC Innovation Center, Stuttgart, Germany
[24] New York University, New York, USA
[25] University of Arizona, Biochemistry Department, Arizona, USA
[26] Max Planck Institute for the Structure and Dynamics of Matter, Hamburg, Germany
[27] Max Planck Institute for the Physics of Complex Systems, Dresden, Germany
[28] Citrine Informatics, Inc., Redwood City, CA, USA
[29] Department of Inorganic Chemistry, Fritz-Haber-Institut der Max-Planck-Gesellschaft, Berlin, Germany
[30] Ecole Polytechnique Fédérale de Lausanne, School of Engineering, Lausanne, Switzerland
[31] Max Planck Institute for Informatics, Saarbrücken, Germany
[32] Department of Statistical Sciences, University of Toronto, Toronto, Canada
[33] Oak Ridge National Laboratory, Oak Ridge, TN, USA
[34] Lead author of the roadmap and to whom correspondence should be addressed scheffler@fhi-berlin.mpg.de

**Abstract**

Science is and always has been based on data, but the terms 'data-centric' and the '4th paradigm' of materials research indicate a radical change in how information is retrieved, handled and research is performed. It signifies a transformative shift towards managing vast data collections, digital repositories, and innovative data analytics methods. The integration of Artificial Intelligence (AI) and its subset Machine Learning (ML), has become pivotal in addressing all these challenges. This Roadmap on Data-Centric Materials Science explores fundamental concepts and methodologies, illustrating diverse applications in electronic-structure theory, soft matter theory, microstructure research, and experimental techniques like photoemission, atom probe tomography, and electron microscopy.

While the roadmap delves into specific areas within the broad interdisciplinary field of materials science, the provided examples elucidate key concepts applicable to a wider range of topics. The discussed instances offer insights into addressing the multifaceted challenges encountered in contemporary materials research.

# Contents

# Section 1: Introduction

The editors: Peter Benner[1], Dierk Raabe[2], Jan-Michael Rost[3], Matthias Scheffler[4], and Gerhard Weikum[5]

[1] Max Planck Institute for Dynamics of Complex Technical Systems, Magdeburg, Germany
[2] Max-Planck-Institut für Eisenforschung GmbH, Düsseldorf, Germany
[3] Max Planck Institute for the Physics of Complex Systems, Dresden, Germany
[4] The NOMAD Laboratory at the FHI of the Max-Planck-Gesellschaft and IRIS-Adlershof of the Humboldt-Universität zu Berlin, Berlin, Germany
[5] Max Planck Institute for Informatics, Saarbrücken, Germany

**Introduction**

Materials science and engineering play a pivotal role in fostering prosperity, enhancing lifestyle, and advancing the development of environmentally sustainable technologies. The field is profoundly interdisciplinary, encompassing physics, chemistry, biology, mathematics, and computer science. It addresses intriguing inquiries such as: Are new semiconductors with increased efficiencies for solar modules available, and can they surpass the flexibility of materials under discussion today? Which catalyst materials would be optimal for a specific chemical reaction, e.g., splitting of water to produce hydrogen? What combination of alloying constituents imparts unique bending strength, extreme hardness, and corrosion-resistant properties of metallic alloys? Furthermore, how should a surface be coated to attain the utmost thermal protection, e.g., for improving the energy efficiency of turbines?

In recent years, materials science has entered an era marked by an unprecedented surge in data, stemming from both experiments and computations. This influx has surpassed the capacities of traditional methods to manage these data effectively. The so-called 4 V challenge is clearly becoming eminent. It can be summarized as follows:

**Volume**: Addressing strategies to manage large datasets efficiently, exploring data storage solutions, and leveraging scalable technologies to handle voluminous data.

**Variety**: Discussing approaches to handle the diverse forms and meanings of data, including data normalization techniques and methods for dealing with heterogeneous datasets.

**Velocity**: Examining ways to cope with the rapid changes in data and the arrival of new datasets in real-time, emphasizing the importance of agile methodologies.

**Veracity**: Exploring methods to assess and enhance the quality and reliability of data, including data validation techniques, quality control measures, and uncertainty quantification.

Amidst these challenges, and most importantly, big data in materials science unveils extraordinary opportunities to advance scientific knowledge and to address important   challenges like those noted above. To seize these opportunities, researchers must adopt fresh perspectives, innovative concepts, and novel methods. This paradigm shift, i.e., a new way of thinking, is commonly referred to as the 4th paradigm of materials research, a term made known by Jim Gray in his inspiring, final talk in 2007 [1]. In essence, 'data-centric research' and the '4th research paradigm' represent a departure from traditional research methodologies. It emphasizes the significance of correlations and statistical predictions, focusing on mean prediction values and variance (or uncertainty) as key elements in the investigative process. In this way the high intricacy of several co- and counter-acting processes is considered. It reflects that big data reveal correlations and dependencies that cannot be seen when studying small data sets, and, in difference to the past, it is accepted that a detailed causal explanation is not always possible. Causal inference, when possible, may not necessarily be expressed in terms of a simple, closed analytic equation or an insightful, simple physical model. We will get back to this point below.

Let us briefly recall the first three research paradigms. Experimental research, the initial paradigm, dates back to the Stone Age and developed first metallurgical techniques in the Copper and Bronze Ages. The control of fire marked a significant breakthrough. In the late 16th century, analytical equations became the central instrument for describing physical relationships, establishing theoretical physics as the second paradigm. The change was led by Brahe, Galileo, Kepler, and Newton. The next chapter started in the 1950s, when electronic-structure theory for solids [2, 3], the Monte Carlo method [4], and molecular dynamics [5, 6] were introduced. These developments enabled computer-based studies and analyses of thermodynamics and statistical mechanics on the one hand and of quantum mechanical properties of solids and liquids on the other hand. They define the beginning of computational materials science, what is nowadays considered the third paradigm of materials research.

Today, big data and AI revolutionize various aspects of life, including materials science. [1, 7, 8] To navigate this 4th paradigm successfully, researchers must embrace new research concepts, and this Roadmap on Data-Centric Materials Science provides a summary of ideas for exploring the data-centric landscape of materials science and engineering. As materials science is a very broad and interdisciplinary field, only some areas of this landscape can be covered. However, we trust that the addressed examples explicate many of the basic concepts and that they can be helpful also for other topics than those addressed explicitly in the different contributions.

Science is and always has been based on data, but the terms 'data-centric' and the '4th paradigm' of materials research signifies a transformative shift towards retrieving and managing vast data collections, digital repositories, and innovative data analytics methods. The integration of AI and its subset ML has become pivotal in addressing all these challenges. In the data analysis, we are looking for structures and patterns in the data. As mentioned above, materials properties and function are often not just governed by one single process but there are many. Some drive, others just facilitate, and again others hinder the materials property or function of interest. The interplay of various processes is very intricate. In analogy to genes in biology, we discuss elemental materials features (e.g., electronegativity of the atoms that build the material) that correlate with the materials property of interest. The primary features that connect with of a certain materials property or function are called the relevant 'materials genes'. Together with environmental parameters (e.g., temperature), they determine (in a statistical sense) the material's property and function.[9]

In recent years, major advances in ML and computing power, in particular the advance of hardware accelerators like GPUs, have enabled deep neural networks, with billions of trainable parameters, leading to breakthroughs in computer vision and natural language processing. A key strength of deep learning is that it addresses not only the objective for classification, regression or other tasks, but also the learning of how to represent the input data itself. Thus, there is no need for explicit feature modeling: images can be ingested as arrays of pixels, and text documents are simply sequences of tokens. High-level structures in visual or textual contents, like people interacting with objects in a scene or argumentation and sentiments in a conversation, are automatically discovered and latently captured by the deep neural network itself.

Obviously, this predictive methodology of deep learning has potential in many application areas, conceivably including materials science and particularly microscopy images. However, the success of deep learning builds on various assumptions, including the availability of large training data with 'independent and identically distributed' (iid) samples. These assumptions are not easily satisfied for materials data, and feature engineering and physics-based modeling is still indispensable. [e.g. Ref. 10]

At its core, ML operates as an interpolation technique, fitting and connecting the data upon which it is trained, applying regularization (or smoothening) to achieve generalization. The ML model excels in exploiting the data space covered by the training data but exhibits diminished reliability when entering uncharted data realms typically called the out-of-distribution (OOD) regime. When the training data are iid or representative of the full population, extrapolation may work. However, for materials science this requirement is hardly fulfilled, i.e., the data selection is governed by subjective and technical issues, and often it is strongly biased and unbalanced. Still, materials scientists are searching for statistically exceptional situations, and important processes are often triggered by 'rare events' that are not or not well covered by the available data set, or smoothed out by the regularization. [e.g. Ref. 11] This all implies caution when applying ML.

Similar to any scientific theory or model, an AI model possesses a range of applicability,[12] often inadequately defined. Consequently, there is an argument advocating the importance of AI interpretability, as it not only sheds light on the underlying mechanism but also provides some confidence in extrapolations. The contributions by Boley et al. (2.1), Ghiringhelli and Rossi (2.2), and Foppa and Scheffler (2.3) address these issues in more detail.

A special point in materials science is that data is typically not big. This implies that some ML methods are not suitable. In general, standard ML methods need to be used with caution and modification or new concepts have been and still need to be developed. Interestingly, Gaussian Process Regression and Random Forests are still often and helpfully used, but several new concepts were established in recent years, e.g., crystal-graph neural network, message passing and equivariance, subgroup discovery, and SISSO (sure independence screen and sparsifying operator). In particular the latter can deal with correlations between a big (even immense) number of elemental materials features (millions or trillions) and just a few data dozens data points of the property of interest. SISSO derives an analytical equation for describing the materials property and its statistical correlation with the relevant materials genes. The approach as well as recent advancements, implementations, and challenges are described by Yao et al. in contribution (3.1).

When data are scarce, the critical request is, that they must be highly accurate, precise, and well characterized. This is summarized by the request that experimental data must be 'clean', but it is not often achieved in materials science and rarely fulfilled in heterogeneous catalysis. The 'clean-data concept' for experimental studies is described in contribution (3.2) by Trunschke et al. Advancements in obtaining high-quality data from electronic-structure theory are described by Kokott et al. in (3.3). The general challenge to find the best-suited AI method for a certain application is severe, and the reproducibility of published AI studies is often problematic. The NOMAD concept is described in contribution (3.4). A strategy to overcome the bottleneck of scarce data in deep learning is the augmentation of a small, accurate data set by synthetically generated data. This is discussed by Giri et al. in contribution (3.5) and exemplified by generating synthetic Hamilton Matrices for deep learning applied to multiphotoabsorption. Spatiotemporal models like random fields and Gaussian processes have demonstrated promising outcomes in integrating data from multiple sources and guiding scientific discovery in various disciplines. Contribution (3.6) by Xian et al. discusses their application to materials science and hints at further directions to be explored to leverage their full potential in materials discovery. When trying to apply machine learning methods that have already proved successful in "hard matter physics" to soft matter, several technical obstacles need to be overcome, including the intrinsic multi-scale nature of this part of condensed matter. Bereau and Kremer argue that when this can be achieved, it would usher soft matter in a new era, where poor scale separation can be efficiently addressed, and insight will be gained for phenomena that are currently too complex for traditional methods (contribution 3.7). In contribution (3.8), Goyal et al. show that significant computational gains can be achieved in the numerical simulation of microstructure continuum mechanics models when traditional direct numerical simulation is replaced by modern deep-learning based methods when the AI models are informed by physical insight. Digitalizing the entire workflow in data-rich imaging techniques in material science from synthesis, sample preparation, data acquisition and post-processing in an integrated way is the topic of contribution (3.9) by Freysoldt et al. There, it is discussed that machine learning techniques can leverage the data science approach by removing the human inspection as the limiting factor to digest larger and larger amounts of data in order to discover relevant, but possibly rare patterns. Recently, large-language models (LLMs) have also entered the field of materials science. Raabe et al. provide an overview and perspective in contribution (3.10).

Section 4 then addresses several applications of data-centric materials science, typically paired with methodological developments. Experimental methods cover photoemission, electron microscopy, and atom-probe tomography. In contribution (4.1), Purcell et al. consider the role of AI in high-throughput materials discovery using computational workflows while Liebscher et al. as well as Schloz et al. discuss the roadmap to AI and ML driven data analytics in scanning transmission electron microscopy (STEM) in contributions (4.2) and (4.3), respectively.  Atom probe tomography is another imaging-based technology to analyze the composition of materials at the near-atomic scale. Its enhancement using ML is the topic of contribution (4.4) by Li et al. In contribution (4.5), Logsdail et al. investigate the potentials of a data-driven approach for heterogeneous catalysis. Finally, in contribution (4.6), Fratzl discusses recent advancements of x-ray scattering and diffraction for materials at the nanoscale with respect to the retrieval and analytics of large amounts of data.

# Section 2: Data and Uncertainty

## 2.1: From Prediction to Action: Critical Role of Performance Estimation for Machine-Learning-Driven Materials Discovery

Mario Boley[1], Felix Luong[1], Simon Teshuva[1], Daniel F. Schmidt[1], Lucas Foppa[2] and Matthias Scheffler[2]

[1] Monash University, Department of Data Science and AI
[2] The NOMAD Laboratory at the Fritz Haber Institute of the Max-Planck-Gesellschaft and IRIS-Adlershof of the Humboldt-Universität zu Berlin

**Status**

In recent years, the materials science community has established a large-scale infrastructure for data sharing that promises to increase the efficiency of the "data-driven" discovery of novel useful materials [1]. Growing data collections are envisioned to lead to increasingly accurate statistical models for property prediction that can significantly reduce the number of necessary experiments or first principles computations and, thus, substantially improve the cost and time for critical discoveries [2]. Indeed, the combination of public datasets and robust statistical estimation techniques like cross validation (CV) enables a collaborative improvement process ("common task framework" [3, 4]). As a result, there are now models that can predict certain materials properties well *on average* with respect to the same distribution as the training data. Unfortunately, the *in-distribution* expected performance, as estimated by CV, is not directly coupled with the performance for the discovery of novel materials: expected performance fails to capture the model behavior for the very few exceptional materials that one aims to discover, and, fundamentally, in-distribution performance is irrelevant for a discovery process that is designed to generate high-performing materials more frequently than they occur in the initial training data.

Recognizing these issues, the community increasingly focusses on active learning approaches [5] like Bayesian optimization for model-driven blackbox optimization [6] (BBO). These methods manage an iterative modelling and data acquisition process and aim to optimize the cumulative "reward" received for the acquired data points over time, such as the maximum property value discovered so far. This process, illustrated in Figure 1, is enabled by an acquisition function that leverages the predictions of a statistical model together with its uncertainty quantification to effectively manage the underlying trade-off of exploration (learning more about the candidate space) and exploitation (aim to sample high value candidates). This shift to consider actions instead of just predictions constitutes an important step towards accelerated materials discovery, but it reveals shortcomings not only in existing modelling approaches but more fundamentally in the methodological framework used to improve those models. In particular, the inapplicability of established performance estimation frameworks based on pre-generated data renders it extremely costly to conclusively compare and to systematically improve methods.

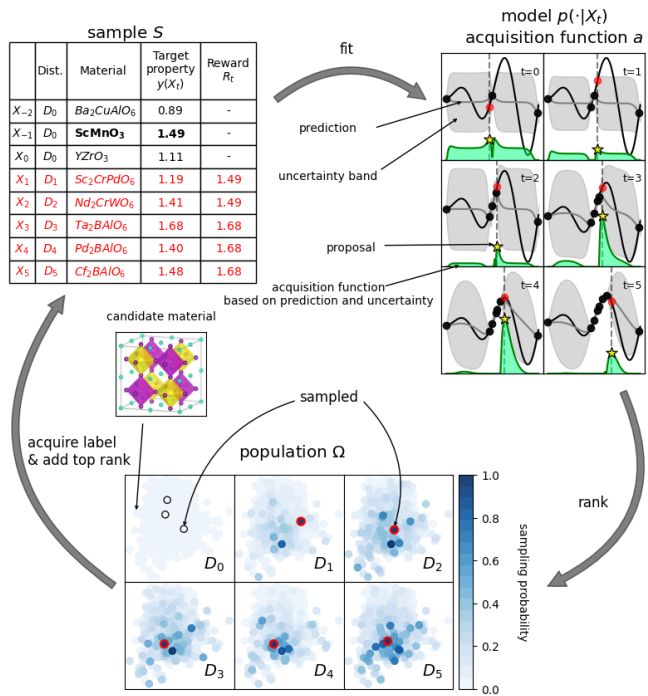

**Figure 1** Schematic steps of iterative data-driven discovery process. At time $t$: (i) probabilistic property model is fitted to sample $\{X_{-N+1}, \dots, X_0; X_1; \dots; X_{t-1}\}$ of materials population $\Omega$, i.e., a conditional density function $p(y \mid x)$ is learned that provides probability density of property value $y$ for material $x$, which gives rise to both (mean) prediction $f(x) = \mathbb{E}_p(Y \mid X = x)$ and uncertainty (variance) $\sigma^2(x) = \mathbb{V}_p(Y \mid X = x)$ where expected value and variance are taken with respect to $p$; (ii) remaining population is ranked by acquisition function, e.g., "expected improvement" of reward $a(x) = \mathbb{E}_p(R_t - R_{t-1} \mid X_t = x)$, which for conditionally normal property models can be computed as $a(x) = f(x) + \sigma^2(x)p(R_t \mid x)/(1 - P(R_t \mid x))$ where $P$ is the modelled cumulative distribution function; and (iii) label for top-ranked material is acquired and added to data sample generating reward, e.g., defined as $R_t = max\{y(X_i) : -N < i \leq t\}$ when maximizing a single property or figure of merit $y$, which incentivizes the discovery of materials with high $y$-value as early as possible in the process. While standard statistical analysis assumes the initial data points $X_{-N+1}, \dots, X_0$ to be drawn with respect to some sampling distribution $D_0$, this distribution does not have to be balanced or representative of the whole population. However, any concentration away from a representative, i.e., uniform, sampling distribution, poses the risk of delayed reward generation, and a misspecified acquisition function or model, in particular one with over-confident predictions, even risks to never escape local maxima represented in the initial data collection. The sampling distribution of subsequent points $D_1, D_2, \dots, D_T$ vary and depend on the combination of model $p$ and acquisition function $a$. Hence, they cannot be pre-generated for new methods rendering label generation a key bottleneck in method development.

## Current and Future Challenges

To illustrate these challenges, let us consider as example the discovery of double perovskite oxides with high *ab initio* computed bulk modulus, where we use two popular statistical models, Gaussian process (GP) regression and random forest (RF), and two BBO data acquisition strategies, *expected improvement* [7] (EI) of rewards and *pure exploitation* [8] (XT). GPs are the traditional BBO model, because their Bayesian approach provides a principled quantification of "epistemic" uncertainty, i.e., uncertainty from a lack of training data related to a specific test point. However, they can struggle already with moderately high-dimensional representations such as the 24 features used in this example. In contrast, RFs are known to work robustly well with high-dimensional feature spaces [9], while their ensemble-based uncertainty quantification does not represent epistemic uncertainty. Interestingly, as shown in Figure 2, CV indicates that RF has the better in-distribution predictive performance not only in terms of squared error but also in terms of log loss, which takes uncertainty into account. Nevertheless, RF is outperformed by GP in terms of the produced discovery rewards, demonstrating that standard in-distribution performance estimation techniques can suggest sub-optimal methods.

This demonstrates that already method selection is a real challenge for practical problems. However, the situation is much worse for methodological research that aims to not only determine, which of a small number of established methods works best, but to test dozens of combinations of models and acquisition functions. Absent innovation in performance estimation, comparing $K$ methods in terms of their expected discovery reward across $L$ repetitons of $T$ rounds requires the acquisition of $KLT$ labels in addition to any pre-generated initial data. This is because, even when starting from a common initial training distribution, each method produces its own sequence of proposal distributions. Since these distributions are unknown *a priori*, there is no way to pre-generate data from them, blocking the usual collaborative improvement process around an initially released dataset. Thus, the prohibitive cost of expected reward estimation currently blocks substantial progress in addressing other important challenges like unsound uncertainty quantification or acquisition function optimization with infinite candidate populations particularly when using non-invertible materials representations.

**Advances in Science and Technology to Meet Challenges**
Given these considerations, a central research goal should be to find reliable approaches for estimating a method's expected discovery reward based on existing data. A simple but infeasible state-of-the-art strategy is to run a method repeatedly using sub-samples of size $n$ from the given dataset as initial data and the sub-sample complement as candidate pool, such that the ratio $n/N$ is close to $N/M$ where $M$ is the overall population size. That is, one naively uses the initial dataset as proxy for the population. For at least two reasons, this simplistic approach is likely to produce misleading results (see Figure 2, middle left). Firstly, the real rewards are determined by the exceptional materials in the tail of the target property distribution, which are almost certainly not well represented in the available dataset. Secondly, changing the absolute sizes of initial data and candidate population misestimates model performance and, more severely, misrepresents the real overwhelming number of uninteresting materials that an efficient search must largely avoid.

Here, we present an adjusted reward estimation approach that provides random initial and candidate sets with realistic absolute numbers of unrepresented exceptional materials as well as distinct ordinary materials to distract from them. Let $X_{(1)}, \ldots, X_{(N)}$ denote the initial data elements in increasing order of their target property or figure of merit values. Based on an estimate $\hat{\alpha}$ of the unrepresented fraction of top materials $\alpha = \#\{X \in \Omega : y(X) \geq y(X_{(N)})\}/M$ create:

1. **an initial dataset** by drawing a size-$N$ bootstrap sub-sample [10], i.e., sample with replacement, from the low property value materials $X_{(1)}, \ldots, X_{(N-\lceil \hat{\alpha} M \rceil - 1)}$ and
2. **a candidate set** consisting of the held-out top $\lceil \hat{\alpha} M \rceil$ materials and an up-sampled and stochastically perturbed set $\tilde{X}_1, \ldots, \tilde{X}_{M - \lceil \hat{\alpha} M \rceil}$ from the unsampled elements of the bootstrap sample.

As shown in Figure 2 (bottom left), reward estimation with this approach performs much better than naïve estimation for our bulk modulus example. It accurately predicts GP with EI to produce the highest bulk modulus and highest cumulative reward out of the four candidate methods. As desired, this is based entirely on the initially available data without requiring the over thousand additional calculations that were needed to confirm this result.

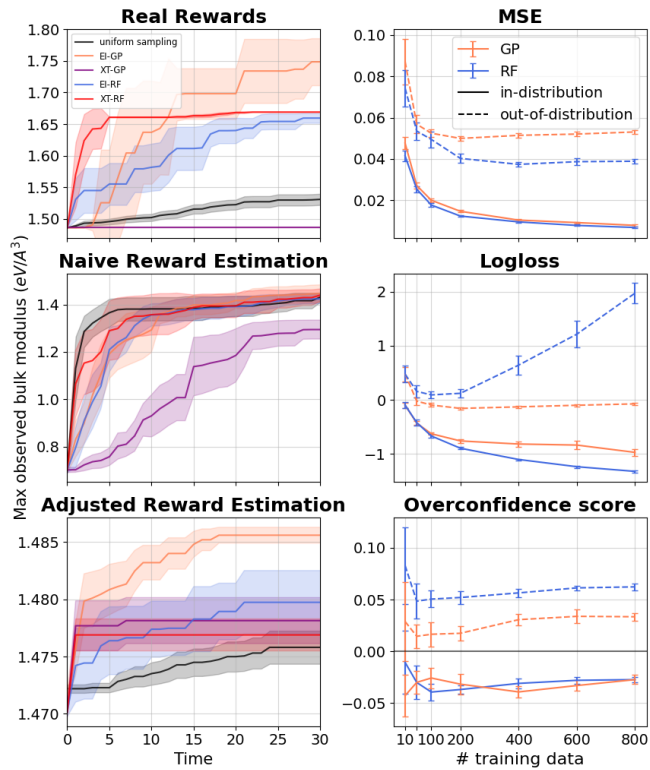

**Figure 2** Performance of Gaussian process (GP) and random forest (RF) models for discovering double perovskites with high bulk modulus. **Left column:** Rewards generated by models with either expected improvement (EI) or pure exploitation (XT) acquisition function as well as their naïve and adjusted reward estimation. Real rewards are mean rewards based on five (EI-GP), nine (EI-RF), ten (both XT), or 100 (uniform) simulations. Estimated rewards are the mean of 20 sub-sampling repetitions. All error bars correspond to 90% confidence intervals. GP with EI has the highest mean reward ($1.66\ eV/A^3$) and discovers the highest bulk modulus ($1.75\ eV/A^3$ on average) in 35 rounds, which is qualitatively predicted by adjusted reward estimation. **Right column:** Model predictive performance estimates in terms of the mean squared error MSE $\mathbb{E}_D(y(X) - f(X))^2$ where $f(X)$ is the prediction for random input point $X$ with property value $y(X)$, log loss $\mathbb{E}_D(log\ p(f(X)\mid X))$ where $p$ is the modelled density of $y(X)$, and overconfidence score $\mathbb{E}(|y(X) - f(X)| - \sigma(X))$ where $\sigma(X)$ where is the modelled standard deviation of $y(X)$ given $X$. Here, all expected values refer to unknown true distributions estimated via 10 (over-confidence score) and 20 (MSE and log loss) repetitions of sub-sampling with replacement from available data (i.e., bootstrap sampling). In-distribution performance is performance with respect to the initial sampling distribution $D_0$, out-of-distribution is with respect to the uniform mixture of the distributions $D_1$ to $D_{100}$ of the data points examined by the discovery process. While RF provides a better mean squared error, both in- and out-of-distribution, its out-of-distribution log loss is increasing with the size of the training, indicating a failure of its uncertainty quantification.

## Concluding Remarks

The lack of reliable approaches to estimate expected discovery rewards from a given dataset is a serious roadblock for the development of active learning methods for materials discovery. Without such estimators, the evaluation of each candidate method requires the acquisition of a potentially large number of labels in addition to any initially available data collection, preventing the usual collaborative process that led to fast-paced improvements of predictive model performance with fixed distributions.

Naïve reward estimation from the initial data typically fails because of unsuitable data proportions and underrepresented extreme events. We presented an adjusted approach that, by correcting for these factors, successfully assesses which combination of acquisition function and statistical model works best for the exemplary task of double perovskite bulk modulus optimization. This or similar approaches could become efficiently computable proxies for real method performances and thus enable fast community-driven improvements to data-driven methods for materials discovery.

**Acknowledgements**
*This work was supported by the Australian Research Council (DP210100045) and the ERC Advanced Grant TEC1p (European Research Council, Grant Agreement No. 740233).*

## 2.2: Reliable Quantification of Uncertainties: The Biggest Challenge for Data-Centric Materials Modeling?

Luca M. Ghiringhelli[1,2] and Mariana Rossi[3]

[1] Department of Materials Science and Engineering, Friedrich-Alexander Universität, Erlangen-Nürnberg, Germany
[2] Physics Department and IRIS-Adlershof, Humboldt Universität zu Berlin, Berlin, Germany
[3] MPI for the Structure and Dynamics of Matter, Hamburg, Germany

**Status**

Artificial-intelligence (AI) and, in particular, machine-learning (ML) modelling is substantially increasing the reach and predictive power of material-science simulations. Such strategies are adopted for two broad classes of applications: a) surrogate modelling of materials properties, e.g., learning energies and forces of given atomic configurations, where the Hamiltonian is known but computationally intensive to evaluate (Refs. 1 and 2 and references therein), and b) materials genomics, i.e., the identification of the features that can explain and be used to model certain materials' property (the genes for that material and property), together with fitting of a predictive model for the given property as function of the identified genes (Refs. 3 and references therein).

Often, the performance of predictive models is focused on averages (e.g., the mean absolute error), and little attention is given to the distribution of errors (e.g, via the so-called violin plots) and to the inspection of the outliers, i.e., the data points that yield the largest prediction errors. Are these data points simply wrongly measured or could they herald some different physical mechanism that was not captured by the model trained to yield slightly acceptable average errors?

Scientifically, it is equally important for a ML model to yield predicted values for new data points and, concurrently, provide reliable uncertainty quantification (UQ). In other words, the model should be able to recognize if it can make a confident prediction solely from the input representation of a test data point, identifying whether it is similar to the data points used for training (interpolatory regime) or dissimilar (extrapolatory regime). The correct metric for assessing this similarity is, however, most often unknown and systematically finding it for a given ML model is one of the most difficult steps for a reliable uncertainty estimate.

Several strategies have been developed for UQ, spanning from rigorous and computationally extremely expensive Bayesian estimates to pragmatic ensemble-of-models training [4-6]. However, many such estimates have been shown to be overconfident when test data are drawn far from the sampling distribution of the training data [7-9]. This limitation represents a serious drawback for the overall reliability of ML models in atomistic simulations, where they promise to deliver first-principles quality results.

**Current and Future Challenges**

Besides the obvious intrinsic benefit of reliably quantifying the uncertainty of an ML model, these estimates are also a vital part of the so-called active-learning (AL) algorithms. AL denotes a strategy where the model constructs new (training) data points either in regions where a property of interest needs to be optimized (exploitation task) or in regions where the model uncertainty is large (exploration task), resulting in a more accurate model with a lower amount of training points. In material science, these algorithms are often desirable, because little initial information is known about a material or materials class and calculating labels (properties) is expensive.

In view of the exploitation task, it is desirable to adopt model classes that allow for a computationally inexpensive optimization (e.g., Gaussian processes). However, the biggest challenge in both surrogate modelling and materials genomics is the UQ in extrapolative regions for the exploration task. In practice, recognizing that a data point belongs to the extrapolation region is the actual conundrum. Statistics and information-theory modelling approaches rely on the fact that training data are representative of the overall population where predictions will be made. In both surrogate modelling and materials genomics applications, the unseen data may carry physical information that is not present in the model training. Electronic-structure data carries a further challenge due to its intrinsic aleatoric uncertainty stemming from numerical convergence and basis sets. It is often difficult, but necessary, to separate it from the model (epistemic) uncertainty, for defining whether training data refinement is needed or whether the model can be really improved.

As for any physical modeling, one does not expect a model to be predictive outside its physical scope. Yet, in the traditional development of physical theories (sometimes referred to as "model-based", as opposed to "data-centric", approach) *describing* the limit of validity of a theory is an essential part of it. Such limits of validity are typically expressed as inequalities as function of key parameters governing the physical property or process. We identify the data-centric identification of the limits of validity of an ML model as, arguably, the biggest challenge in AI applied to materials science.

## Advances in Science and Technology to Meet Challenges

The full acceptance of ML tools within the community, for both surrogate modeling and materials genomics, may depend on two interrelated aspects: The introduction of algorithms for a) reliable UQ, especially for data points that are outside the training distribution and b) finding *explanations* why any given outlier is an outlier.

For the first aspect, in the realm of surrogate model potentials, Bayesian-based frameworks offer an intrinsic definition of uncertainty, which can be judiciously used [11]. For neural-network architectures, committee ensemble models can deliver some degree of uncertainty prediction. In both cases, correctly accounting for correlations in the training set data is essential for avoiding overconfident model predictions [12], but UQ can still be unreliable for out-of-sample data points. A promising alternative is the use of deep ensembles or variations thereof. Finally, because the surrogate model is trained to predict energy and forces, but these quantities are almost never the observable that is being sought in a simulation, advances in error definition and propagation through derived properties have been gaining much attention [13].

For the second aspect, a promising route is the use of subgroup discovery for the identification of the so-called domains of applicability (DAs, regions of the input space where a predictive model yields small errors) [10], which are given in form of descriptive rules, i.e., inequalities over a set of features, identified among a larger set of candidates. Although it has been shown that DAs can be found and the descriptive rules give insight on the analyzed ML models, the method has not been yet further developed to systematically identify outliers and exploited to improve the underlying ML model, e.g., in an AL fashion.

## Concluding Remarks

The recent literature has shown that, with carefully selected training data sets and physical expertise (domain knowledge), the resulting ML predictive models allow for important discoveries in materials science. However, unleashing the full potential of data-centric approaches and fulfilling their promise to deliver results of ab initio quality requires that the uncertainty of the predictions be quantified.  This UQ needs to be robust and reliable, and the related algorithm should be relatively straightforward to implement, such that users have a transparent access to it.

Although reliability has to be prioritised, any UQ algorithm must not add a substantial computational cost to the ML model it is being applied to, since in materials modelling efficiency is often a core requirement to achieve meaningful simulations. This observation applies both to the realm of surrogate modelling where, e.g., millions of force evaluations with uncertainty quantification need to be carried out, and to the realm of materials genomics where, e.g., millions of candidate systems need to be classified including this quantification. Achieving such a framework requires the community to adopt more widespread standards and work together on benchmarking efforts targeted at error prediction.

Reaching this goal would enable the systematic, fully data-centric improvement of the learned model, via the active-learning strategies, and the assessment of the limits of validity of the learned models.

## Acknowledgements
LMG acknowledges funding from the NOMAD Center of Excellence (European Union's Horizon 2020 research and innovation program, grant agreement № 951786) and the project FAIRmat (FAIR Data Infrastructure for Condensed-Matter Physics and the Chemical Physics of Solids, German Research Foundation, project № 460197019). We acknowledge financial support from BiGmax, the Max Planck Society's Research Network on Big-Data-Driven Materials-Science. We thank Matthias Scheffler for a critical read of the earlier version of this manuscript.

## 2.3: Towards a Multi-Objective Optimization of Subgroups for the Discovery of Materials with Exceptional Performance

Lucas Foppa[1] and Matthias Scheffler[1]

[1] The NOMAD Laboratory at the Fritz Haber Institute of the Max-Planck-Gesellschaft and IRIS-Adlershof of the Humboldt-Universität zu Berlin, Berlin, Germany

**Status**

Artificial-intelligence (AI) approaches in materials science usually attempt a description of all possible scenarios with a single, global model. However, the materials that are useful for a given application, which requires a special and high performance, are often statistically exceptional. For instance, one might be interested in identifying exceedingly hard materials, or materials with band gap within a narrow range of values. Global models of materials' properties and functions are designed to perform well in average for the majority of (uninteresting) compounds. Thus, AI might well overlook the useful materials. In contrast, subgroup discovery (SGD) [1,2] identifies *local* descriptions of the materials space, accepting that a global model might be inaccurate or inappropriate to capture the useful materials subspace. Indeed, different mechanisms may govern the materials' performance across the immense materials space and SGD can focus on the mechanism(s) that result in exceptional performance.

The SGD analysis is based on a dataset $\tilde{P}$, which contains a known set of materials. $\tilde{P}$ is part of a larger space of possible materials, the full, typically infinite population $P$. For the materials in $\tilde{P}$, we know a target of interest $Y$ (metric or categorical), such as a materials' property, as well as many candidate descriptive parameters $\varphi$ possibly correlated with the underlying phenomena governing $Y$ (Fig. 1). From this dataset, SGD generates propositions $\pi$ about the descriptive parameters, e.g., inequalities constraining their values, and then identifies selectors $\sigma$, conjunctions of $\pi$, that result in SGs that maximize a quality function $Q$:

$$Q(\tilde{P}, SG) = \left(\frac{s_{SG}}{s_{\tilde{P}}}\right)^{\gamma} * \left(u(SG, \tilde{P})\right)^{1-\gamma} . \text{(Eq. 1)}$$

In Eq. 1, the ratio $s_{SG}/s_{\tilde{P}}$ is called the coverage, where $s_{SG}$ and $s_{\tilde{P}}$ are the number of data points in the SG and in $\tilde{P}$, respectively. The utility function $u(SG, \tilde{P})$ measures how exceptional the SGs are compared to $\tilde{P}$ based on the distributions of $Y$ values in the SG and in $\tilde{P}$. $Q$ establishes a tradeoff between the coverage (generality) and the utility (exceptionality), which can be tuned by a tradeoff parameter $\gamma$. Typically, the identified selectors only depend on few of the initially offered candidate descriptive parameters. The identified SG selectors (or rules) describe the local behaviour in the SG and they can be exploited for the identification of new materials in $P$.

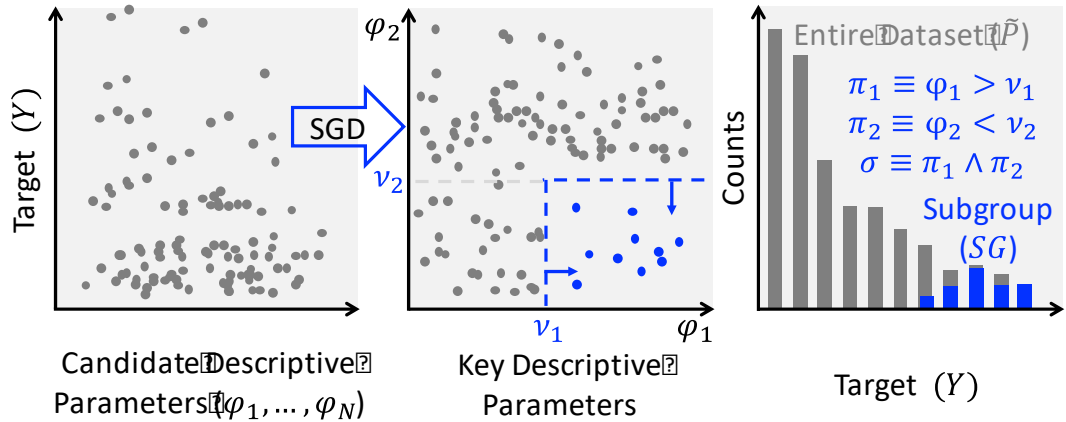

**Figure 1** Subgroup discovery (SGD) identifies descriptions of exceptional subselections of the dataset. These descriptions (rules) are selectors $\sigma$ constructed as conjunctions of propositions $\pi$ about the data. The symbol $\wedge$ denotes the "AND" operator.

## Current and Future Challenges

The potential of SGD to uncover local patterns in materials science has been demonstrated by the identification of structure-property relationships, [3] and by the discovery of materials for heterogeneous catalysis. [4] Additionally, using (prediction) errors as target in SGD, we identified descriptions of the regions of the materials space in which (machine-learning) models have low [5] or high errors. [6] Thus, the domain of applicability (DoA) of the models could be established. Despite these encouraging results, the advancement of the SGD approach in materials science requires addressing key challenges:

- The quality function introduces one generality-exceptionality tradeoff, among a multitude of possible tradeoffs that can be relevant for a given application and that can be obtained with different $\gamma$. For instance, the required hardness of a material depends on the type of device in which it will be used and the DoA of a model depends on the accuracy that is acceptable to describe a certain property or phenomenon. However, choosing the appropriate $\gamma$ and assessing the similarity - or redundancy - among the multiple rules obtained with different tradeoffs are challenging tasks.

- Widely used utility functions assess the exceptionality of SGs by comparing the data distribution of the SG and that of $\tilde{P}$ via a single summary-statistics value. For example, the positive-mean-shift utility function for metric target favors the identification of SGs with high $Y$ values only based on the means of the two distributions. Thus, it is often assumed that the distributions are well characterized by the chosen summary-statistics value and that $\tilde{P}$ is representative of the full population $P$. However, distributions in materials science are typically non-normal and $\tilde{P}$ might not reflect the infinitely larger, unknown $P$. This calls for the consideration of utility functions that circumvent the mentioned assumptions.

- The mechanisms governing materials can be highly intricate and the relevant descriptive parameters to describe a certain materials' property are often unknown. Thus, one would like to offer many possibly relevant candidate parameters and let the SGD analysis identify the key ones. However, optimizing the quality function is a combinatorial problem with respect to the number of descriptive parameters and efficient search algorithms are therefore crucial. [7]

## Advances in Science and Technology to Meet Challenges

In order to address some of these open questions, we approach the SGD as a multi-objective-optimization problem for the systematic identification of SG rules that correspond to a multitude of generality-exceptionality trade-offs. Coherent collections of SG rules are obtained by considering the Pareto front of optimal SGD solutions with respect to the objectives coverage and utility function, as illustrated for the example of identification of perovskites with high bulk moduli in Fig. 2. Once the coherent collections of SG rules are identified, the overlap between SG elements can be used to assess their similarity. A high similarity between SG rules might indicate that the rules are redundant. Thus, the similarity analysis can be used to choose the SG rules that should be considered for further investigation or exploitation.

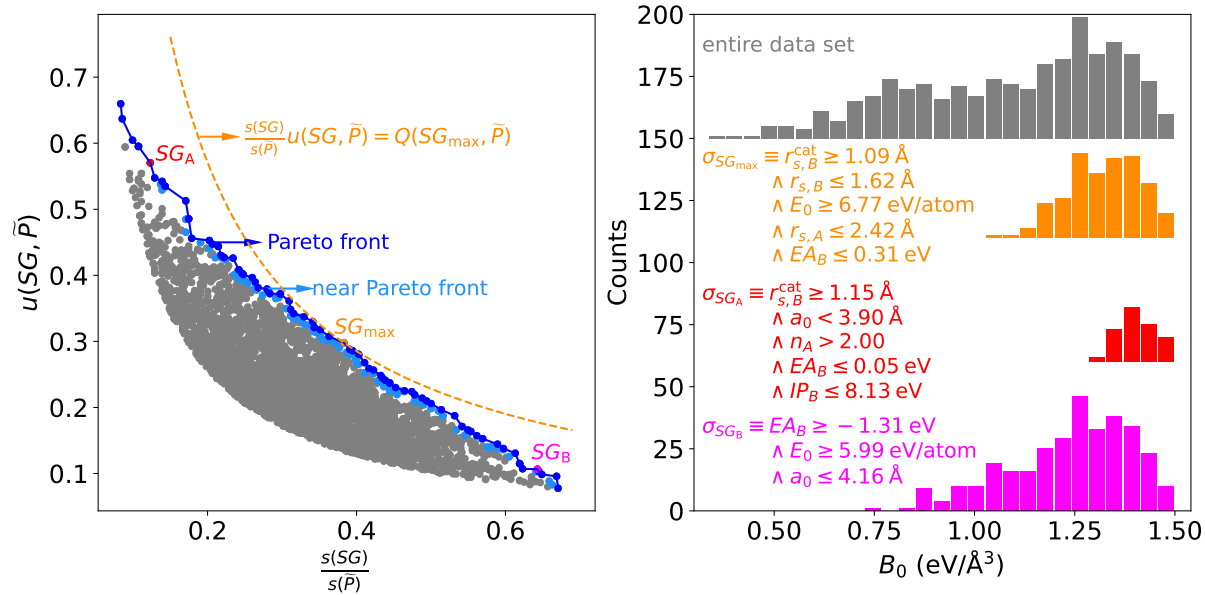

**Figure 2** Left panel: A coherent collection of SG rules describing $ABO_3$ perovskites with high bulk modulus ($B_0$) is identified at the Pareto front of SGD solutions with respect to the objectives coverage and the utility function *cumulative Jensen-Shannon divergence*. Right panel: The identified rules constrain the values of the radiii of the $s$ orbitals of isolated $A$, $B$ and $B^{+1}$ species ($r_{s,A}$, $r_{s,B}$ and $r_{s,B}^{cat}$, respectively), the electron affinity and ionization potential of isolated $B$ species ($EA_B$ and $IP_B$, respectively), the expected oxidation state of $A$ ($n_A$), the equilibrium lattice constant ($a_0$), and the cohesive energy ($E_0$).

Noteworthy, the *cumulative Jensen-Shannon divergence* ($D_{JS}$) [8] between the distribution of bulk moduli in the SG and in the entire dataset is used as quality function in the example of Fig. 2. $D_{JS}$ assumes small values for similar distributions and increases as the distribution of target values in the SG is, e.g., shifted or narrower with respect to the distribution of the entire dataset. Crucially, $D_{JS}$ does not assume that one single summary-statistics value represents the distributions. Divergence-based utility functions addressing, e.g., high or low target values, will thus be an important advance. We note that the utility function might also incorporate information on multiple targets or physical constraints that are specific to the scientific question being addressed. [9] However, in order to ensure that the training data is representative of the relevant materials space one would like to cover, the iterative incorporation of new data points and training of SGD rules in an active-learning fashion might be required.

**Concluding Remarks**

SGD can accelerate the identification of exceptional materials that may be overlooked by global AI models because it focuses on local descriptions. However, further developments are required in order to translate the SGD concept to the typical scenario of materials science, where datasets might be unbalanced, or not be representative of the whole materials space and the most important descriptive parameters are unknown. The multi-objective perspective introduced in this contribution provides an efficient framework for dealing with the compromise between generality and exceptionality in SGD. The combination of this strategy with efficient algorithms for SG search and with a systematic incorporation of new data points to better cover the materials space will further advance the AI-driven discovery of materials.

**Acknowledgements**

This work was funded by the NOMAD Center of Excellence (European Union's Horizon 2020 research and innovation program, Grant Agreement No. 951786) and by the ERC Advanced Grant TEC1p (European Research Council, Grant Agreement No. 740233).

# Section 3: Methods

## 3.1: Building Portable Artificial-Intelligence Software for the Exascale

Yi Yao[1], Thomas A. R. Purcell[1,2], Sebastian Eibl[3], Markus Rampp[3], Luca M. Ghiringhelli[1,4], and Matthias Scheffler[1]

[1] The NOMAD Laboratory at the FHI of the Max-Planck-Gesellschaft and IRIS-Adlershof of the Humboldt-Universität zu Berlin, Berlin, Germany
[2] University of Arizona, Biochemistry Department, Arizona, USA
[3] Max Planck Computing and Data Facility, Garching, Germany
[4] Department of Materials Science and Engineering, Friedrich-Alexander Universität, Erlangen-Nürnberg, Germany

**Status**
Modern high-performance computing (HPC) systems are evolving towards greater heterogeneity and diversification. The heterogeneity is due to the use of specialized processing units for specific tasks, nowadays with a strong (commercial) focus on AI-specific algorithms. This strategy, led by companies like Nvidia with their (general-purpose) GPUs and tools like CUDA, is driven by the need to enhance computational performance while containing electrical-power consumption and total cost. Present-day exascale and pre-exascale systems commonly integrate GPUs with CPUs of different architectures and vendors. Additionally, alternative accelerators like tensor processing units (TPU), neural processing units (NPU), field programmable gate arrays (FPGA), and emerging technologies like neuromorphic and quantum processors add to the array of high-performance computing options. These will further contribute to the heterogeneity and diversification of high-performance computing but have not yet broken into scientific computing. Except for the quantum processor, the other accelerators adhere to classical architectures characterised by varying levels of parallelism.

To tap the power of accelerators, AI codes must incorporate efficient internode communication schemes (like the well-established MPI) and align with programming models associated with the available accelerators. Examples include CUDA for Nvidia GPUs, ROCm for AMD GPUs, or SYCL/DPC++ for Intel GPUs. The neural network-based AI codes often rely on the availability and development of frameworks such as pyTorch and tensorflow, where the developers of these frameworks take the burden to adapt the framework to accelerators. For instance, pytorch provides versions of its framework with support for CUDA or ROCm backends. However, not all of the AI methods can be seamlessly translated into a neural-network representation and not all applications are well suited for neural networks. Consequently, significant adaptation is required, leading to limited accelerator support. For example, the widely used decision-tree-based AI library, XGBoost, offers a CUDA version, but is still lacking a ROCm equivalent.

**Current and Future Challenges**
The current challenge involves developing performance-portable and maintainable code for AI methods, in general, on HPC systems. This task will become even more challenging with the increasing heterogeneity of HPC systems. In a typical HPC system, internode communication is necessary and the message passing interface (MPI) has proven to be a flexible and effective solution for doing this. The so-called MPI+X paradigm combines MPI with intra-node parallelization models and/or accelerator offloading models (X). The choice of accelerator offloading model is largely determined by the specific accelerators in use, together with problem requirements and personal taste.

We summarise various strategies that have been developed to address this challenge in Figure 1. For offloading work onto an accelerator, the most straight-forward approach would be to write the algorithms with accelerator-specific interfaces such as CUDA. While this in principle allows to tap all (performance) capabilities, these interfaces are limited to specific accelerators. Given the abundance of existing CUDA code in scientific computing, AMD and Intel have introduced tools to facilitate the translation of such code into their HIP/ROCm and SYCL/DPC++ language, which more or less resemble the semantics of CUDA. Moreover, the HIP and SYCL programming models even claim some universality by enabling code execution not only on AMD or Intel GPUs, respectively. However, the viability and broader adoption of these comparably recent approaches remains to be demonstrated.

An alternative approach involves the utilisation of architecture-independent - and typically more abstract - programming models, which come in various forms. One category employs compiler directives to manage loop parallelization and data management. Examples include OpenMP [6] and OpenACC [7]. Programming with these directives aims at a single codebase compatible with different accelerators. Directive-based approaches can also facilitate the reuse of existing CPU-based code and enable an incremental code-porting workflow by successively "offloading" performance-critical parts of the code. Success-stories have been observed adapting these models [8].

Another approach are C++ portability frameworks such as Kokkos[1] and RAJA[5]. They provide high-level parallel abstractions such as the parallel implementation of the traditional "for", "reduce", and "scan" operations, which the framework maps to specific hardware backends that use the corresponding platform-native programming models. These may, in addition, serve as forerunners for corresponding extensions to be added to the C++ standard.

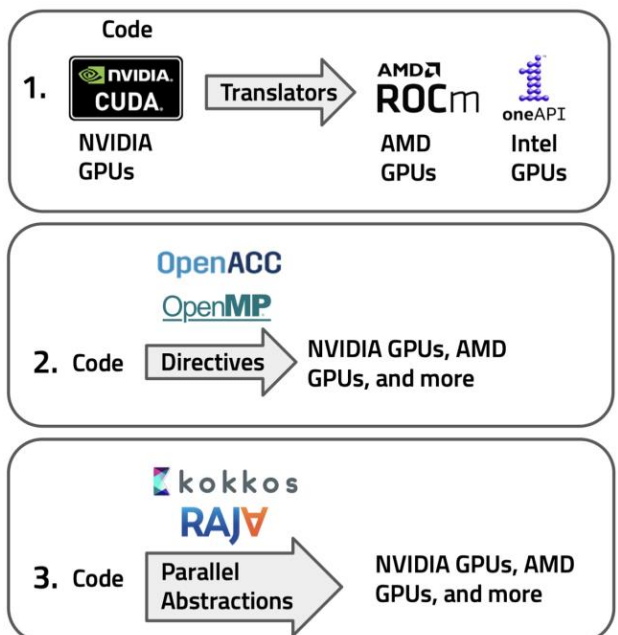

**Figure 1** Strategies to port codes onto diversified and heterogeneous high-performance computers. Translators convert from one programming model to another, directives are compiler instructions to dictate how a piece of code

should be compiled, and parallel abstractions define how a computation workload may be calculated in parallel. The library will then map the abstractions to GPUs.

**Advances in Science and Technology to Meet Challenges**

As an example of how the code-portability challenge can be met for an originally developed AI application which is different from deep-learning, we outline the porting of an implementation of the sure-independence screening and sparsifying operator (SISSO) to GPUs using Kokkos. SISSO is a combination of symbolic regression and compressed sensing. It first generates a list of up to trillions of analytical expressions from an initial set of primary features and mathematical operators. It then uses an $\ell_0$ regularised least squares regression to find the best low-dimensional linear model from the generated expressions. In preparation for (pre-)exascale computing, we converted the most computationally intensive components of SISSO, i.e. expression generation and $\ell_0$ regularisation, in our initial MPI+OpenMP code to a MPI+OpenMP+Kokkos implementation, in order to demonstrate scalability and portability on exascale-ready HPC platforms.

Throughout the development, we refactored the data structures to suit the access pattern of accelerators, and carefully optimised the memory migration between host and device. This results in an approximately tenfold speedup by the GPUs of two generations of Nvidia GPUs for a test problem with ~60 billion generated features and ~36 billion least squares regression problems (see Figure 2). The code also scales to at least 64 nodes, see Figure 2. We expect that scaling to much higher node counts can be achieved with increasing the size of the training dataset. Notably, the same code also runs on AMD Instinct MI200 GPUs with a similar speedup without any code modifications, except for compilation settings. Given that the Kokkos framework supports backends for CUDA, HIP, SYCL, OpenMP, we expect our code can also be smoothly ported to other accelerators. Since Kokkos is developed and maintained with strong commitments by the US DOE laboratories, we expect it to receive continuous support and will extend to future  HPC hardware.

One key question when using an abstraction framework is how close its performance comes to the "native", i.e. architecture-specific programming models. In our case, we compared the performance of our batched least-squares-regression algorithm (for $\ell_0$ regularization) to a native CUDA implementation co-developed with Nvidia engineers. This new CUDA code is about twice as fast as the Kokkos version. However, it is worth noting that Kokkos' continuous development is promising. For instance, during our development, transitioning from Kokkos version 3 to version 4 resulted in a 10% speedup without requiring any code modifications from us in the application code.

## Porting Process

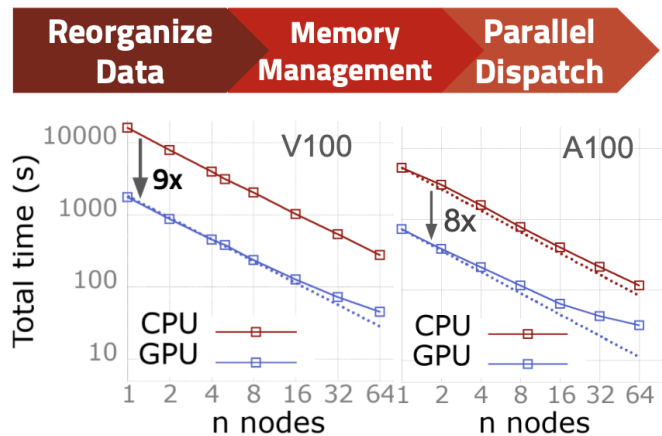

**Figure 2** The process and performance of porting SISSO++ on different HPC platforms with Kokkos library. The scaling test is performed on (1), the Talos cluster at MPCDF up to 64 nodes with 40 Intel Skylake CPU cores and 2 Nvidia V100 GPUs on each node, and (2), the Raven cluster at MPCDF, on up to 64 nodes with 72 Intel Xeon Icelake CPU cores and 4 Nvidia A100 GPUs on each node.

## Concluding Remarks

The growing diversity and heterogeneity in (pre-)exascale high-performance computing poses significant challenges to software developers, including performance portability and code maintainability. To tackle these issues, developers have adopted various strategies, such as code duplication (typically abstracted internally by some application-specific interfaces), (semi-)automatic code translation, directive-based portability models, and high-level abstraction frameworks. For our SISSO++ code, which is an AI application not readily amenable to the well-established (and portable) AI frameworks like, e.g., Tensorflow, we opted for the MPI+X paradigm which is well established in HPC, specifically using MPI+OpenMP+Kokkos. The usage of the Kokkos abstraction framework enhances both the code performance and portability, and it also helps reduce code-maintenance burdens. The Kokkos framework is also expected to pave the way for adopting the parallel abstract concepts in future C++ language standards. Due to the generality of the Kokkos framework, and already proven for SISSO++ by a seamless transition over two generations of Nvidia GPUs, we anticipate that our SISSO++ code will easily adapt also to future HPC architectures. Our porting strategy outlined here can serve as an example for other non-neural network based AI code development efforts.

## Acknowledgements

We thank Patrick Atkinson and Markus Hrywniak from Nvidia for their support in optimising the CUDA-native reference implementation for our benchmark. The project received financial support from BiGmax, the Max Planck Society's Research Network on Big-Data-Driven Materials Science, the NOMAD Center of Excellence (European Union's Horizon 2020 research and innovation program, Grant Agreement No. 951786) and the ERC Advanced Grant TEC1p (European Research Council, Grant Agreement No. 740233).

## 3.2: Clean-Data Concept for Experimental Studies

Annette Trunschke[1], Lucas Foppa[2] and Matthias Scheffler[2]

[1] Department of Inorganic Chemistry, Fritz-Haber-Institut der Max-Planck-Gesellschaft, Berlin, Germany
[2] The NOMAD Laboratory at the Fritz Haber Institute of the Max-Planck-Gesellschaft and IRIS-Adlershof of the Humboldt-Universität zu Berlin, Berlin, Germany

**Status**

Materials design typically targets an application that requires the synthesis of a material which is characterised by measurable and reliable properties and functions that are maintained during its use. Inexpensive and abundant raw materials, reproducibility, and scalability are decisive factors for success. The relationships between the structure and the function of a material are usually complex and intricate and they prevent a strictly *in-silico* design for realistic conditions. Thus, the experimental input is crucial. Artificial-intelligence (AI) methods have the potential to reduce the significant efforts related to the synthesis and characterization of materials, accelerating materials discovery. However, rigorously conducted experiments that provide consistent training data for AI are indispensable. They directly determine the reliability of generated insights.

The applications of AI in materials science are diverse.[1-2] For example, the optimization of synthesis and functional properties in high-throughput experiments requires mathematical models which are iteratively trained in order to ensure an efficient experimental design. The elucidation of materials structures can be facilitated by AI. Besides, new materials can be predicted via the identification of correlations and patterns in experimental and computational data sets. This leads to a variety of data set structures. The interdisciplinary nature of materials science and the multitude of experimental techniques applied produce a broad spectrum of data formats, all of which can ultimately be traced back to spectroscopic, thermodynamic or kinetic relationships and are already standardised to some extent. Experimental data in materials research are usually not "big data", which places additional demands on the methods of data analysis.

However, if the data becomes FAIR,[3] *i.e.*, Findable, Accessible, Interoperable, and Reusable and open, *i.e.*, generally accessible after publication, machines can systematically analyse this information beyond the boundaries of a single laboratory and field of research, learn from it and develop disruptive solutions.[4] A particularly sustainable generation of insight is achieved through the use of interpretable AI algorithms that uncover descriptors, *i.e.*, correlations between key physical parameters and the material properties and functions.

**Current and Future Challenges**

Predictions could be more reliable if the materials function of interest was determined exclusively by the bulk properties of the material. However, when the material's function is affected, or even governed by interfacial and kinetic phenomena, such as in case of batteries, sensors, biophysical applications or catalysis, the relationships between the materials parameters and the function become extremely complex. On the one hand, this is caused by the strong influence of defects and minor impurities. On the other hand, the material properties respond to the fluctuating chemical potential of the environment in which they are used. This gives metadata such as the sequence of experimental steps and the time frame a particular importance.[5]

In order to make experimental data useful for a digital analysis, the measurements have to follow so-called "standard operating procedures" (SOPs), as is already common practice in some research areas. An important cornerstone for such workflows is the introduction of certified standards that enables the direct evaluation of measured data when they are published together with the results of the standard. Awareness of the need for rigorous work and standardization of experiments has grown considerably in academic research in recent times and it is reflected in initiatives for standardized measurement procedures and test protocols (Reference 4 and references cited therein).

The currently most common form of publication in scientific journals does not support the direct electronic access to the data. The use of natural language processing (NLP) tools is one approach to analyse and understand human language in published articles.[6] These computer science techniques can help to identify trends, but do not provide consistent data sets, as data in publications are not presented uniformly, for example often only in the form of graphical representations, and data as well as metadata are not necessarily provided completely.

The most effective solution to enable the use of experimental data in AI is to apply machine-readable SOPs in automated experiments. In this way, standardized and complete data and metadata sets can be generated that can be shared after publication in repositories, as is already widely done in computational materials science and synthesis of complex molecules. The latter also requires the development of ontologies. We note that digital SOPs are an important preliminary step for enabling autonomous research by robots in the future.[7]

**Advances in Science and Technology to Meet Challenges**
The most common AI methods require large amounts of data, and only the smallest part of available data in materials science meets the quality requirements for data-efficient AI. In a use case study,[8] we have shown how a "clean" data-centric approach in interfacial catalysis enables the identification of descriptors based on a data set that can be generated in the experimental practice with reasonable effort (Figure 1). Here the term "clean data" refers to the fact that the considered materials were carefully synthesized, characterized, and tested in catalysis according to SOPs reported in an experimental handbook.[5]

Large-scale applications in the field of energy storage such as water splitting and the efficient use of resources in the production of consumer goods are generally based on highly complex catalysed reactions at interfaces. The selective oxidation of the short-chain alkanes ethane, propane and *n*-butane to valuable olefins and oxygenates was chosen as an example of a reaction type that is known for its complicated reaction networks. Control over the selective formation of desired products in this network and the minimization of $CO_2$ formation requires sophisticated catalyst materials and adapted reaction conditions.

Experimental procedures that capture the kinetics of the formation of the active phase from the catalyst precursors have been designed and specified in a SOP.[5] A typical set of twelve chemically and structurally diverse catalyst materials was included in the study that combines rigorously conducted clean experiments in catalyst synthesis, physicochemical characterization and kinetic evaluation with interpretable AI using the sure-independence-screening-and-sparsifying-operator (SISSO) symbolic-regression approach.[9] Previously obtained empirical findings are correctly reflected by the data analysis, which proves the value of the data set.

Interpretable AI goes far beyond empirical interpretations. It addresses the full complexity of the dynamically changing material and the full catalytic process by identifying non-linear property-function relationships described by mathematical equations in which the target catalytic parameters depend on several key physicochemical parameters of the materials measured *in operando* and after different stages in the life cycle of the catalyst. These key descriptive parameters, that the AI approach identifies out of many offered ones, reflect the processes triggering, favouring or hindering the catalytic performance. In analogy to genes in biology, these parameters are called "materials genes" of heterogeneous catalysis, since they describe the catalyst function similarly as genes in biology relate, for instance, to the color of the eyes or to health issues. Thus, these materials genes capture complex relationships. They describe a correlation (with uncertainties) but they do not provide the detailed description of the underlying processes.

This data-centric approach discloses which of the often time-consuming and expensive characterization techniques are really important for the catalyst design. The chemist is also provided with practical guidelines for optimizing certain materials properties to further improve the catalyst's function.

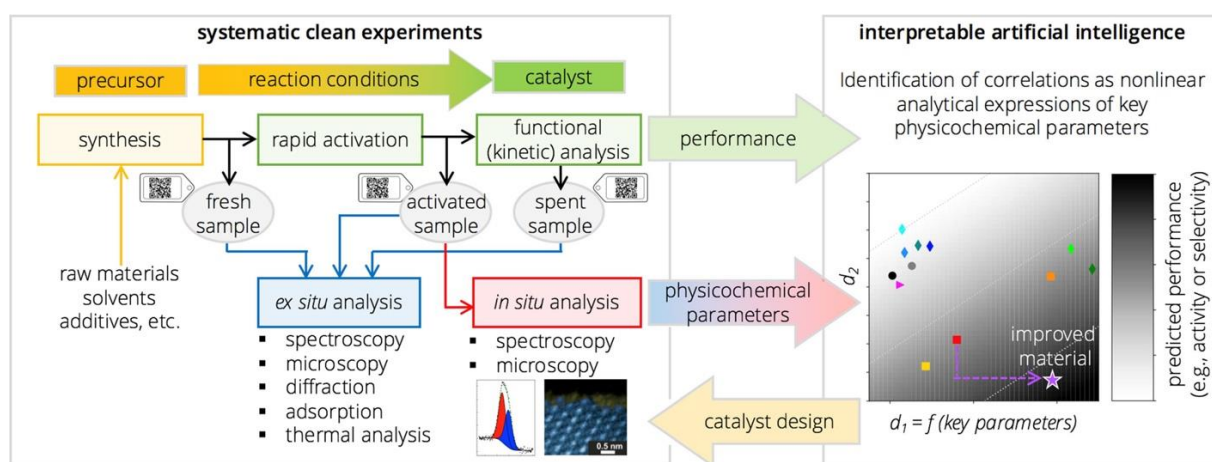

**Figure 1** Clean experiments designed to capture the kinetics of the formation of the catalyst active state under reaction conditions are used to generate a consistent and detailed data set, which is then analysed via the sure-independence-screening-and-sparsifying-operator (SISSO) artificial-intelligence (AI) approach in order to uncover the key physicochemical parameters describing the catalytic performance. The figure is reproduced from reference 8 (ACS, CC-BY 4.0).

**Concluding Remarks**

Reproducibility is probably the most basic and crucial requirement of materials science. AI is an efficient tool in materials research and development, but its application requires that we change the way we work and deal with data. Complete, uniform and reliable data sets are required that comply with the FAIR principles. These can be obtained by working across laboratories according to standard operating procedures ("handbooks"), which also include the analysis of benchmarks. Important elements for the gradual development of autonomous materials research,[10] in addition to technical progress in robotics, are the use of machine-readable handbooks, automated experiments with standardized data analysis and upload to local data infrastructures as well as the standardized publication of experimental data in overarching open repositories.

**Acknowledgements**

This project received financial support from BiGmax, the Max Planck Society's Research Network on Big-Data-Driven Materials Science,  the NOMAD Center of Excellence (European Union's Horizon 2020 research and innovation program, Grant Agreement No. 951786) and the ERC Advanced Grant TEC1p (European Research Council, Grant Agreement No. 740233). The use case study "Clean data in oxidation catalysis" was conducted in the framework of the BasCat - UniCat BASF JointLab collaboration between BASF SE, Technical University Berlin, Fritz-Haber-Institut (FHI) der Max-Planck-Gesellschaft, and the clusters of excellence "Unified Concepts in Catalysis"/"Unifying Systems in Catalysis" (UniCat https://www.unicat.tu-berlin.de/ UniSysCat https://www.unisyscat.de).  Funding by the Deutsche Forschungsgemeinschaft (DFG, German Research Foundation) under Germany's Excellence Strategy - EXC2008 - 390540038 - UniSysCat is acknowledged. Work on overarching repositories is funded by the Deutsche Forschungsgemeinschaft (DFG, German Research Foundation), in the framework of the project FAIRmat—FAIR Data Infrastructure for Condensed Matter Physics and the Chemical Physics of Solids, project number 460197019. Studies regarding the automation of experiments in catalysis research are funded by the Federal Ministry of Education and Research (BMBF) in the framework of the CatLab project, FKZ 03EW0015B.

### 3.3: Towards Efficient and Accurate Input for Data-Driven Materials Science from Large-Scale All-Electron DFT Simulations

Sebastian Kokott[1,2], Andreas Marek[3], Florian Merz[4], Petr Karpov[3], Christian Carbogno[1], Mariana Rossi[5], Markus Rampp[3], Volker Blum[6] and Matthias Scheffler[1]

[1] The NOMAD Laboratory at the FHI of the Max-Planck-Gesellschaft and IRIS-Adlershof of the Humboldt-Universität zu Berlin, Berlin, Germany
[2] Molecular Simulations from First Principles e.V., Berlin, Germany
[3] Max Planck Computing and Data Facility, Garching, Germany
[4] Lenovo HPC Innovation Center, Stuttgart, Germany
[5] MPI for the Structure and Dynamics of Matter, Hamburg, Germany
[6] Thomas Lord Department of Mechanical Engineering and Materials Science, Duke University, Durham, USA

**Status**

The quality of input data is critical for data driven science. Detailed, high-level (i..e, quantum many-body theory based) simulations, although expensive, can provide immensely valuable data on which other methods can build, if three main issues can be addressed: First, the system size of accurate quantum-mechanical simulations is often restricted by the computational complexity of the underlying simulation algorithms. Second, the accuracy of the predicted data for new complex materials critically depends on the accuracy of the specific physical model chosen to derive quantum-mechanical simulation data, limiting subsequent data-driven models. Third, the number of atomic-scale configurations that must be covered for a statistically sound description grows dramatically with the complexity of a material, necessitating more and faster high-level calculations to provide input data for subsequent, AI-driven research. Simulations of real-world materials require addressing all three points at the same time.

Hybrid density functionals (hybrids) have emerged as a practical reference method for ab initio electronic-structure-based simulations because they resolve several known accuracy issues of lower levels of density-functional theory (DFT) while offering affordable computational cost on current high-performance computers. There are two main computational bottlenecks for atomistic simulations using hybrid DFT: evaluating the non-local exact exchange contribution and finding the solution of a generalized eigenvalue problem (matrix diagonalization). Here, we discuss advances and perspectives for both challenges as recently implemented in the all-electron code FHI-aims [1].

The current reach of these methods is documented by run times and scaling of hybrid DFT simulations for several challenging materials, including hybrid organic/inorganic perovskites [2] and organic crystals, with up to 30,000 atoms (50,000 electron pairs) in the simulation cell. Despite such large systems sizes, the simulations can be run with moderate computational resources.

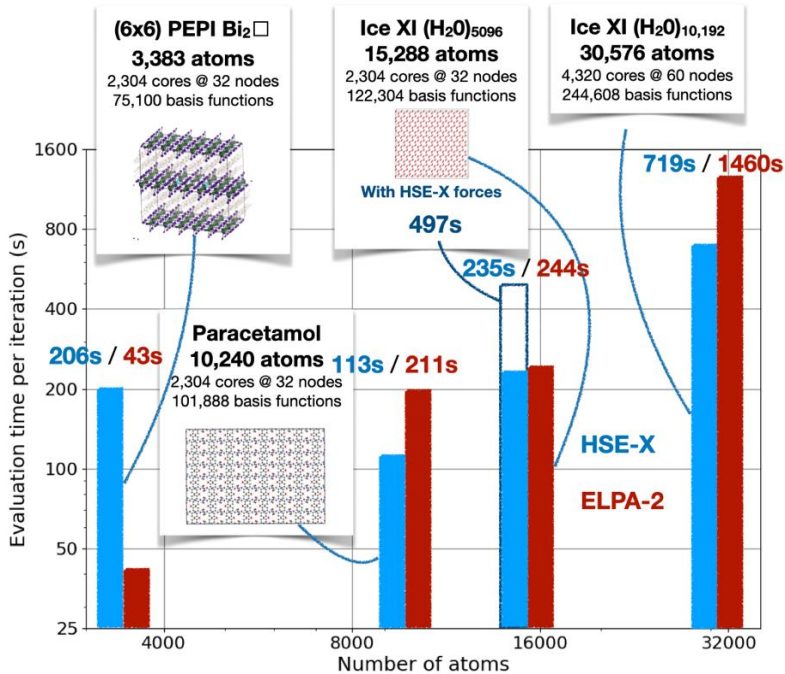

**Figure 3** Average runtime to evaluate the exchange operator (blue bars) and the ELPA eigenvalue solver (red bars) per self-consistent field iteration. The HSE06 hybrid functional was used for all simulations. The following systems were simulated (from left to right): phenylethylammonium lead iodide (PEPI) with a defect complex [2], a 4x4x4 paracetamol supercell, a 15,288-atoms Ice XI supercell (including a force evaluation), and a 30,576-atoms Ice XI supercell. All calculations were carried out on the Raven HPC cluster at the MPCDF using Intel Xeon IceLake (Platinum 8360Y) nodes with 72 cores per node.

## Current and Future Challenges

A resolution-of-identity-based real-space implementation of the exact exchange algorithm [3–5] was optimized to allow for much improved exploitation of sparsity and load balancing across ten thousands of parallel computational tasks. Results show drastically improved memory and runtime performance, scalability, and workload distribution on CPU clusters. The improvements pushed the simulation limits beyond 10,000 atoms, compared to an earlier implementation that reached system sizes around 1,000 heavy atoms [4]. This new code implementation can perform computation of energy, forces, and stress for periodic and non-periodic systems for several fashions of hybrid density functionals. In addition, for materials including heavy elements, perturbative spin-orbit coupling can be combined with the hybrids [6]. Due to the inherent $O(N^3)$ scaling, the solution of the eigenvalue problem beyond 10,000 atoms becomes the bottleneck during the simulations.

The direct eigensolver library ELPA [7] has long offered unrivalled performance for parallel matrix diagonalizations. Extensive profiling, fine tuning and work on portability was carried out to adapt ELPA to the most current HPC architectures, further reducing the time for the diagonalization bottleneck for simulation sizes up to many thousands of atoms. Key to future success of ELPA is exploiting full capabilities of GPU-accelerated high-performance clusters. ELPA already has a well-established support for NVIDIA GPUs [8,9]. Recently, we ported ELPA to AMD GPUs, enabling the solution of a problem with a matrix size with leading dimension of more than 3 million on 1024 AMD-GPU nodes of the LUMI pre-exascale system at CSC in Finland.

Although the library APIs for AMD and NVIDIA are very similar, we find very different run-time and performance behaviour for the ELPA code. Thus, a new abstraction layer driving the GPU computations within ELPA has been implemented. Below this abstraction layer, the vendor specific implementations co-exist and can be independently developed and optimized. We believe that this very flexible approach facilitates the integration of upcoming new architectures, e.g., Intel GPUs.

Similar GPU strategies will be needed for the exact exchange algorithm, but are not yet exploited, as the porting of CPU code to GPU architecture is not at all straightforward. In the CPU implementation, the inherent sparsity of real-space approach keeps the size of matrices used for dense matrix-matrix operations moderate. Thus, with the current algorithm the full capabilities of GPUs cannot be used, and speedups would be limited by communication. An overhaul of the algorithm, and GPU-specific storage and communication patterns will be needed to make it amenable for heterogeneous, GPU-accelerated architectures.

### Advances in Science and Technology to Meet Challenges

The achievements for hybrid DFT simulations demonstrated above is a big success and paves the way to efficient use of exa-scale resources in the future. Still, the accuracy of hybrids is limited by construction. The required fraction of exact exchange is an open point. A related question is the treatment of the electron correlation – hybrid density functionals addressing this point only insufficiently. Approaches using range-dependent parameters for the fraction of exact exchange or double hybrids are a way forward to improve the accuracy of the ab initio model. The GW approach and the CCSD method provide much more accurate access to electronic structure quantities per se, but the complexity of these methods will limit their application to smaller system sizes for the foreseeable future.

From a technological point of view, we think that sufficiently large memory per node and task will be needed for any enhanced electronic structure method, i.e., usually non-local operators are evaluated, which require finding a good balance between communication across nodes and tasks and storing data. Here, the tighter integration of accelerators within the HPC node, as, for example, expected for the upcoming Nvidia (Grace-Hopper) and AMD (MI300) technologies, looks very promising. There are two main hurdles for scientific software developers: library APIs for solving mathematical and physical problems are partially vendor-specific and/or not performance optimal. Addressing both points increase the reach of scientific code (and in turn reduces the need for code duplication) and will reduce the overall cost of research significantly. As a difficult task remains the optimization of communication patterns between CPUs and GPUs for specific architectures. Also new workload distribution models might be needed to better use all available resources, e.g., compute with GPUs and CPUs at the same time (right now often CPUs are idling while GPUs do the work).

### Concluding Remarks

The new exact exchange algorithm implemented in FHI-aims and the highly optimized ELPA library enables simulations of large system sizes at moderate runtimes. On the one hand, these implementations allow one to increase statistical sampling to address the huge configuration space that comes with the large system sizes. On the other hand, the accuracy of hybrid DFT simulations is sufficient for many applications. We believe that with the aid of future exa-scale resources in combination with sophisticated data-driven models, hybrid functionals will be established as default method for DFT simulations of materials. In general, exploiting sparsity is key to low-scaling electronic structure methods for large scale simulations. Real-space algorithms using localized wavefunctions are especially well suited. Nevertheless, the data distribution and communication pattern may need architecture-specific optimizations that complicates software design and code maintenance.

**Acknowledgements**

This project received financial support from BiGmax, the Max Planck Society's Research Network on Big-Data-Driven Materials Science, the NOMAD Center of Excellence (European Union's Horizon 2020 research and innovation program, Grant Agreement No. 951786) and the ERC Advanced Grant TEC1p (European Research Council, Grant Agreement No. 740233).

### 3.4: Choosing AI Analysis Tools and Enacting their Reproducibility: The NOMAD AI Toolkit

Luca M. Ghiringhelli[1,2,] Luigi Sbailò[1], Ádám Fekete[1], Markus Scheidgen[1], and Matthias Scheffler[3]

[1] Physics Department and IRIS-Adlershof, Humboldt Universität zu Berlin, Berlin, Germany
[2] Department of Materials Science and Engineering, Friedrich-Alexander Universität, Erlangen-Nürnberg, Germany
[3] The NOMAD Laboratory at the FHI of the Max-Planck-Gesellschaft and IRIS-Adlershof of the Humboldt-Universität zu Berlin, Berlin, Germany

**Status**

When, at the end of 2014, the NOMAD Repository & Archive [1, 2] went online, it was the first data infrastructure in computational materials science that fulfilled what was later and independently defined by the acronym FAIR (Findable, Accessible, Interoperable, and Reusable). This definition and the request that scientific data should be FAIR was introduced in a very general scientific-data context by Wilkinson *et al.* in 2016 [3]. As of today, the NOMAD Repository stores input and output files from more than 50 different atomistic (*ab initio* and molecular mechanics) codes and totals more than 13 million entries, uploaded by over 500 international authors from their local storage, or from other public databases. The NOMAD Archive stores the same information, but converted, normalized, and characterized by means of a metadata schema, the NOMAD Metainfo [4], which allows for the labeling of most of the raw data in a code-independent representation. One of the benefits of normalized data is that they are accessible in a format that makes them suitable for direct artificial-intelligence (AI) analysis.

NOMAD also offers the AI toolkit [5], a JupyterHub-based platform for running notebooks on NOMAD servers, without the need for any registration or downloaded software. The data-science community has introduced several platforms for performing AI-based analysis of scientific data, typically by providing rich libraries for AI. General-purpose frameworks such as Binder [6] and Google Colab [7], as well as materials-science dedicated frameworks such as pyIron [8] , AiidaLab[9] , and MatBench [10] are the most used by the community. In all these cases, a big effort is devoted to education via online and in-person tutorials. The main specificity of the NOMAD AI toolkit is its connection with the extensive NOMAD Archive. Moreover, together with the NOMAD Oasis [2], users can work with their private as well as community data within the same software platform and using the same API.

**Current and Future Challenges**

Besides providing the framework for performing custom-made AI analysis, the NOMAD AI toolkit provides a set of tutorial notebooks introducing users step by step into both the most popular and widely known AI methodologies, with showcase applications in materials science, and into the more advanced ones, i.e., methodologies that have been published in the latest years. Due to the very nature of the Jupyter technology, these tutorial notebooks are interactive, in the sense that users can modify lines of codes and check the effect of the modifications. Also, the tutorial notebooks have direct access to the whole NOMAD data, so that users can apply the learned techniques to new data, including data uploaded by them.

Importantly, the AI toolkit includes notebooks that present actual AI software as used for producing results for peer-reviewed publications. This feature suggests that scientific *reproducibility* can reach its full potential, at least for AI analysis tools. For instance, users can re-train AI models with exactly the same set of hyperparameters as used in the original publications, on exactly the same data, including the train/validation/test set splits. A piece of information that nowadays is not required in peer-reviewed publications. However, such addition would be scientifically appropriate as it would directly enable the reproducibility of reported results. The NOMAD AI toolkit enables this important step.

As already noted by the proponents of the FAIR principles for scientific software [11], providing complete information on the algorithms and software used to analyze data is all but trivial. This is particularly challenging if one wants to provide live software that can be run on demand, mainly because pieces of software, e.g Python scripts for an AI analysis, require a virtual environment where libraries that are used for efficiently performing certain routine tasks are installed. These libraries get repeatedly updated, and unfortunately backward compatibility is not necessarily ensured. This means that the same set of commands that at release time allows to install and run a software, at a later point in time may not yield a correct installation. Besides, in the case that the software is run in a container (as for the NOMAD AI toolkit), when a new container is created the software for the container platform gets updated. In other words, special care and planning has to be devoted to maintaining the whole ecosystem of software so that exactly the same datasets yield in time exactly the same AI models and therefore exactly the same predictions over the same test data.

**Advances in Science and Technology to Meet Challenges**
Platforms like the NOMAD AI toolkit also foreshadow the scientific-reproducibility utopia. Much of the technology that allows for reaching these goals still needs to be developed, but some important steps have been taken already. First of all, Jupyter notebooks can be uploaded to NOMAD as easily as the data. The upload timestamps and other *provenance* metadata that allow for the unique identification of each analysis script. Furthermore, users are encouraged to provide a rich set of metadata that are made searchable and therefore will allow other users to locate the notebooks by e.g., model class for the AI analysis, or by used libraries, including their versions. In its current state, the NOMAD AI toolkit allows for the findability and interoperability of the AI-analysis software. In fact, a unique container is currently used for all the notebooks, thus allowing for a full interoperability among the different AI analysis tools. The complexity of the maintenance of such an environment rapidly increases with the number of uploaded notebooks which poses challenges to ensure  that stored notebooks can run over the years and produce the same results. However, each set of obtained results, including all the intermediate results along the analysis workflow, can be stored (according to FAIR principles) and automatic tests could be run to check the conformity of the results produced by the re-trained model with the reference ones. Knowing that some piece of code is at some point in time unable to reproduce old results is the necessary condition to try and fix the code in order to conform with the reference results. This solution, which requires quite some human effort, introduces a possibly interesting generalization on the idea of reproducibility, which in some sense is a black-box requirement. I.e, in each step of the analysis, the same input needs to yield the same output, but the details inside the black box are allowed to change.
A radically alternative route is to partly renounce to a full interoperability among the notebooks and maintain several different containers within the NOMAD AI toolkit. Such an approach would allow for the creation of specific containers that are not updated, thus allowing for the software installed therein to be always executable. Although the tools used in these not-updated containers cannot always be combined with software installed into other containers , it can still be deployed on new data that have been uploaded at a later time.

**Concluding Remarks**
The introduction and gradual implementation of the FAIR practices for scientific-data management and stewardship revealed that another crucial component of scientific research needs to adopt the FAIR concepts: The scientific software for data production and analysis. As for data, the key point is the *reproducibility* of research finding, i.e., the practical possibility to re-obtain the same results starting from the same hypotheses (the input settings) and methods.

Clearly, providing only the input data and results in a data archive, even if fully FAIR-data compliant, is not enough for reproducibility, if part of the results are obtained in an incompletely documented way and/or via some custom-tailored analysis software, which is not properly stored and versioned.

The NOMAD AI toolkit already enables re-run AI software on FAIR data for a relatively small set of Jupyter notebooks at the price of human-intensive maintenance. The grand-challenge is to develop a strategy to scale up such maintenance in a (semi-)automatic fashion, so that all AI tools from the community can be preserved according to FAIR practices, fully achieving scientific reproducibility.

Clearly, these reproducibility concepts and the use of Jupyter notebooks also imply that newcomers to AI can use the software that already exists at the NOMAD infrastructure, train themselves and adjust and advance the analysis tools towards their own but different applications.

**Acknowledgements**

We acknowledge inspiring discussions on the future directions of the AI toolkit with Thomas Hammerschmidt, Kevin Jablonka, and José Márquez Prieto.

The project received funding from BiGmax, the Max Planck Society's Research Network on Big-Data-Driven Materials Science, the NOMAD Center of Excellence (European Union's Horizon 2020 research and innovation program, grant agreements № 676580 and 951786), the ERC Advanced Grant TEC1p (European Research Council, grant agreement № 740233), and the project FAIRmat (FAIR-Data Infrastructure for Condensed-Matter Physics and the Chemical Physics of Solids, German Research Foundation, project № 460197019).

### 3.5: Synthetic Hamilton Matrices for Deep Learning

Sajal K. Giri[1], Ulf Saalmann[2], and Jan M. Rost[2]

[1] Department of Chemistry, Northwestern University, Illinois, USA
[2] Max Planck Institute for the Physics of Complex Systems, Dresden, Germany

**Status**

Training of Deep Learning (DL) models requires a large amount of data in the first place and the data set must be sufficiently diverse for the network to be transferable such that it produces unbiased predictions. At the same time, the data size needs to be balanced to compensate for the cost of their generation. Strategies to deal with scarce data problems include Transfer Learning (TL), Self-Supervised Learning, Generative Adversarial Networks (GANs), Model Architecture, Physics-Informed Neural Network, and Deep Synthetic Minority Oversampling techniques to name a few recent approaches, as pointed out in [1].

Here, we focus on a specific route to overcome the scarce training data bottleneck, namely the generation of *random synthetic training data* under suitable constraints determined by the physics involved [2-4]. In our approach we aim at modelling system dynamics by encoding it into a Hamilton matrix for the interaction of (bound) electrons with intense laser light. The latter can be very noisy and fluctuating from shot to shot, as produced by X-ray Free Electron Lasers (XFEL). We vary the elements of the Hamilton matrix randomly about a matrix of an existing model system in one physical dimension (1D), creating synthetic Hamilton matrices (SHMs) for systems which could but do not necessarily exist in nature for which calculations can be done quickly compared to real 3D systems. From the large set of SHMs augmented with different deterministic realizations of noisy laser pulses, we compute photoelectron spectra to train a fully-connected deep neural network (DNN). Figure 1 shows an application of the DNN (trained with spectra from SHMs) to a real 3D system for which it predicts, without knowing the system explicitly, how the noisy spectrum would look like if a "clean" (Gaussian) laser pulse would have been used. The good agreement with the ground truth demonstrates that the trained DNN can be transferred from 1D to 3D problems and gives confidence in our SHM deep learning concept (SHM-DL).

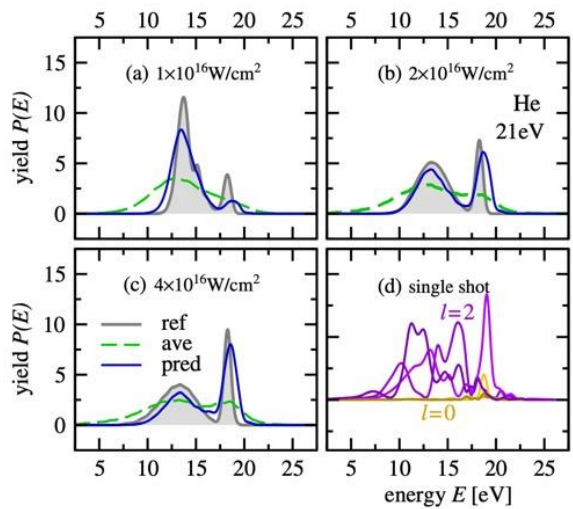

**Figure 1.** Photoelectron spectra for the He atom (a)–(d) from noisy laser pulses with a central photon energy of 21 eV and peak intensity $I$ as given in the panels. In (d) the spectra from different single noisy pulses are shown with dominant contributions from angular momentum *l*=2 (purple) and *l*=0 (yellow) after absorption of two photons by the ionized electron at $I = 2 \times 10^{16}$ W/cm². Panels (a)-(c) show the spectra which result from averaging the single-shot spectra (green-dashed), the reference (i.e. the spectra calculated from a clean Gaussian laser pulse, gray-shaded) and the predicted spectra (blue solid) for a clean pulse by the DNN (Figure with permission from [2]).

Very recently, the idea of synthetic data generation based on existing data has been taken up for composite materials [4], where a limited number of original full-field micro-mechanical simulation data are randomly rotated in physical 3D space to generate additional data to train a recurrent neural network for the non-linear elasto-plastic response of Short Fiber Reinforced Composites. Similar ideas using TL are being explored in other areas [5].

**Current and Future Challenges**

An important problem in the context of spectra generated by XFEL double pulses is the delay between the pulses which jitters in an unknown way from shot to shot. The SHM-DL approach can extract the time-delay of a double-pulse from the spectrum it has generated. Importantly, we can sort single-shot noisy spectra according to the time-delay of the double pulse with which the spectra were generated. With a second network the time-delay sorted pulses, binned over small delay intervals (1fs) can be purified as shown in Figure 2. This constitutes a substantial generalization to predict a hidden parameter (the time-delay of the pulses) [3].

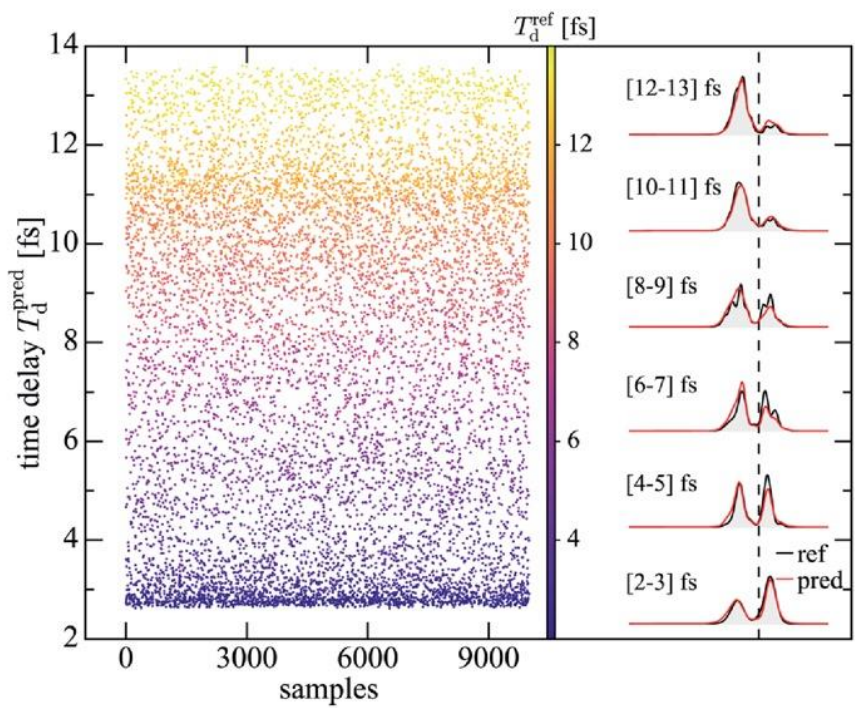

**Figure 2.** Reconstruction of double pulse time-delay and purification of noisy spectra for a single Hamilton matrix taken from test data. Single-shot fluctuating spectra for random time-delays are passed through the trained network to reconstruct the underlying time delays which are shown as scattered points where the color gradient represents the reference time-delay. We consider 12 intervals of time delay in the range 2–14 fs with an interval length of 1 fs. All single-shot spectra which fall into interval of time-delay are averaged. The averaged spectra are passed through another network which maps averaged noisy spectra to purified ones. The predicted purified spectra (red) are compared to reference spectra (black). Adapted with permission from [3].

The task the SHM-DL has successfully completed so far, is the mapping of spectra generated by noisy pulses to spectra generated by Gaussian (Fourier limited) pulses. Can we also predict via SHM-DL maps spectra for other clean pulse forms, e.g., for pulse forms which are not even realizable experimentally? This would be very interesting for systems whose response to light cannot be computed (too complicated) but measured, e.g., with noisy pulses as described, since with SHM-DL we do not need to compute the "true" spectrum of the desired system.

*The primary vision* of the SHM-DL approach is a 21$^{st}$ century spectroscopy. Applied to molecular ro-vibrational spectra, e.g., it could replace the traditional normal mode model for the assignment and classification of spectral, leaving it to the trained network to associate appropriate SHMs with the spectrum, thereby classifying it by means much more flexible than traditional, structurally predefined normal modes.

*The long-term goal* is to develop SHM-DL to become capable of identifying a single SHM (or a small group of SHMs) which describe the system so well, i.e. represent the system, such that from the reconstructed SHM(s) time-dependent system evolution in general and other observables can be computed/predicted. This would constitute a physics-rooted form of generalization which delivers at the same time physical insight as it provides an optimal parametrization of a physical system with a Hamilton matrix of chosen size. First attempts are promising that identify SHMs in relation to two- or multidimensional spectroscopy [6].

**Advances in Science and Technology to Meet Challenges**

Technically, even the SHM-DL approach remains a challenge regarding the computing power needed to numerically produce the spectra (training data) from the SHMs. Hence, (i) a reduction of the required training data size by better knowledge of the underlying physics is desirable as well as (ii) a reduction of computational costs by ultra-efficient quantum propagation in time to obtain the spectra [7]. (iii) Furthermore, the computed spectra as training data must be balanced. For the time being this is done by simply discarding spectra from the training set which are too close to each other. However, this implies a large waste of computing time. To reduce this waste several advances are desirable: Firstly, use of an optimal metric to determine the Euclidean "distance" between two spectra. Here, recently the Wasserstein metric has become popular [8], or approximations to it which are computationally cheaper. A more elegant, physics-oriented advance would be to find an approximate inversion of the SHM-to-spectrum map, or any other way which allows us to shift the balancing of the spectra to suitable choices of the SHMs.

Thinking ahead, the idea of SHMs could be realized not with DNNs but other DL approaches. Most promising are GANs or variants thereof, where the relevant SHM is constructed by the GAN from a random one successively with computationally costs eventually reduced compared to the present SHM-DL sampling approach.  Moreover, the GAN approach would directly predict an SHM which describes the system's coupling to light.

Finally, and almost trivial since true for many DL applications: SHM-DL would benefit enormously from a possibility for inherent error quantification.

## Concluding Remarks

We have introduced the idea of synthetic Hamilton matrices (SHMs), random representations for the dynamics of systems coupled to light, which could exist but not necessarily exist in nature. This approach enables sampling of the training space solving the dynamics with SHMs efficiently incorporating sufficiently generic features to be transferable to real systems. They serve the purpose  to augment training data for DL with DNNs. We demonstrated that this SHM-DL approach works by  purifying photoelectron spectra from noisy pulses and identifying pulse delays which jitter in an unknown manner, as supplied by X-ray Free-electron Lasers. The approach is physics-oriented and therefore promises physics insight beyond the prediction of spectra through the DL based identification of the relevant Hamilton matrix from a spectrum for a system, unknown to the DNN.

### Acknowledgements

The project received funding from BiGmax, the Max Planck Society's Research Network on Big-Data-Driven Materials Science.

## 3.6: Spatiotemporal Models for Data Integration

R. Patrick Xian[1], Jason Hattrick-Simpers[2], Ralph Ernstorfer[3], Stefan Bauer[4]

[1] Department of Statistical Sciences, University of Toronto, Toronto, Canada
[2] Department of Materials Science and Engineering, University of Toronto, Toronto, Canada
[3] Institute for Optics and Atomic Physics, Technical University of Berlin, Berlin, Germany
[4] School of Computation, Information and Technology, Technical University of Munich & Helmholtz AI, Munich, Germany

**Status**

Understanding the structure-property relations of materials, and optimizing chemical synthesis or device manufacturing processes requires integrating multimodal datasets from both theory and experiments [1] that often encompass spatial and temporal dependence. The explicit spatiotemporal characteristics may be exploited in model-building for data integration. Historically, spatial and spatiotemporal models were largely developed in the contexts of geoinformatics, biostatistics, and quantitative ecology, many of which remain underappreciated by the materials science community. In these models, the temporal and spatial subsystems are typically considered in 1D and 2D/3D, respectively. Spatiotemporal models describe their subsystems jointly to capture the interactions through covariance functions or dynamical processes derived from physical knowledge [2]. They are structured and interpretable and are considerably more tractable than first-principles methods.

Random fields (RFs) and Gaussian processes (GPs), also known as kriging, are two established categories of models designed for spatial and spatiotemporal data. RFs already have established use in the statistical modeling of microstructured materials [3], while GPs are invented in mining engineering and they are a classic example of surrogate models. We discuss here three diverse examples from materials data science that indicate their broad applicability and utility. (i) In metal additive manufacturing, Saunders et al. [4] combined three GPs with distinct characteristics to model the pairwise relationships between materials microstructure, melt pool geometry, and mechanical property obtained from multiphysics simulation, all of which are also time-dependent. (ii) In photoemission spectroscopy, Xian and Stimper et al. [5] constructed a Markov RF with nearest-neighbor interaction and transformed the band dispersion reconstruction problem into a classification problem. The coordinates in their problem are the two momenta and energy of photoelectrons, the use of pre-computed energy bands from electronic structure theory provides an effective initialization. (iii) In combinatorial materials screening, Kusne et al. [6] constructed a GP in the chemical compositional space to guide the search for the optimal stoichiometry within a family of tertiary phase-change materials. Their algorithm was integrated into a synchrotron beamline and may be run in a closed loop driven by active learning.

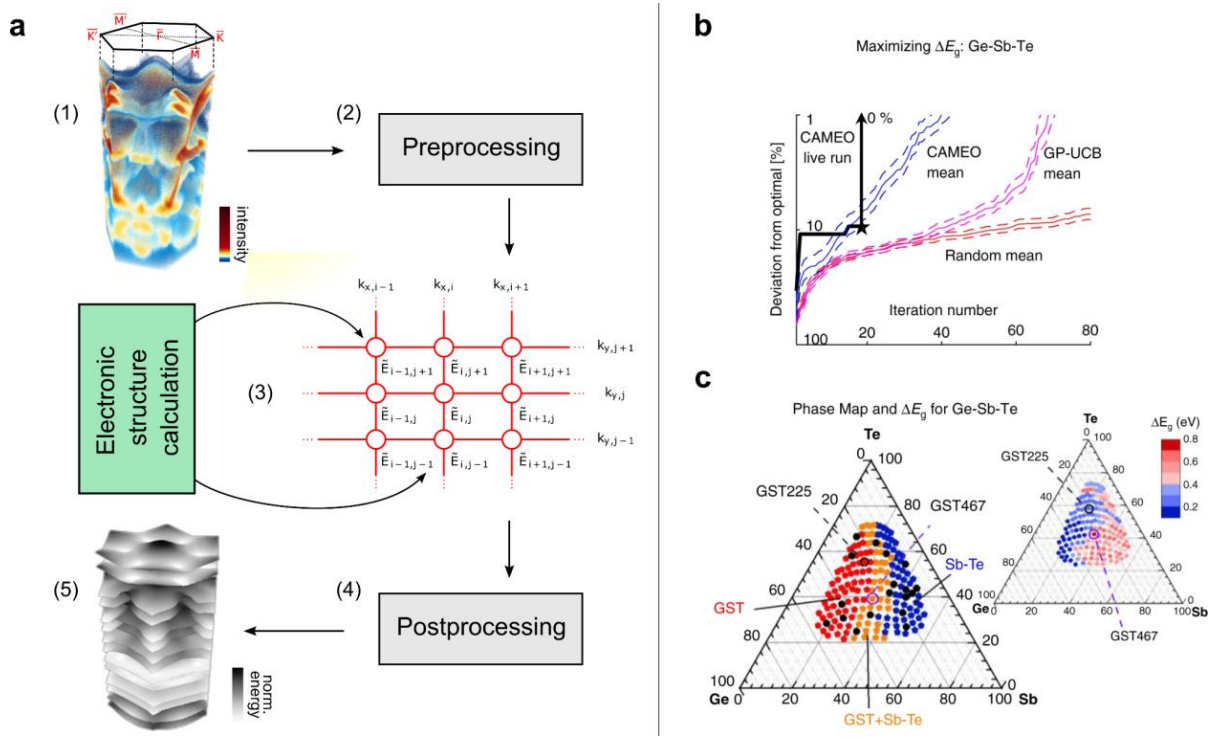

**Figure 1.** Illustrations of spatial models of (left) photoemission data in the energy-momentum coordinates using a Markov random field [5] and (right) combinatorial material screening data in the chemical compositional space using a Gaussian process [6].

**Current and Future Challenges**

One defining characteristic of materials science data is its abundance of data types, from videos to images to atomic structures [1]. Comparatively, spatiotemporal models, besides the classic examples like RFs and GPs, may also take the form of point processes [7], state space models [8], and diffusion processes [9], which are as yet not used for data integration, but have their respective benefits to representing specific data types. Besides coordinates with concrete physical meaning, one could also consider direct spatial or dynamical models of the latent space in data integration, as it is often more robust to noise and dimensional scaling artifacts, especially for multiple data modalities. This leads to the question of problem mapping from data type to model category and subsequent model specification as the primary challenges. The three examples in the previous subsection illustrate that building spatial and spatiotemporal models are not limited to the physical dimensions attached to their original meaning. The straightforward way for problem mapping is to first identify the data types related to a particular problem, then consider the native data types to the model and find the match. For example, point process models would be suitable for modeling the transport of point defects because of their sparse distribution.

Secondly, we should pay special attention to the data quality in the subsystems to be integrated, including resolution, unit size (such as pixel size for image data), missingness, structuredness, and fidelity (such as noise level). Many of these problems are not yet formally addressed, thereby motivating further research on a case-by-case basis guided by domain knowledge. For example, data resolution and fidelity affect the choice of integration ordering, i.e., from high to low or in reverse. For experimental data, the unit size is usually not the same as the resolution because of blurring introduced by the instrument response. Thirdly, we should consider the scalability of the model during development, which may be left unnoticed until later in model deployment using real-world data. For spatiotemporal models that aim to handle large datasets, scalability is often a primary determinant of model choice.

**Advances in Science and Technology to Meet Challenges**

The two main paradigms in materials science that benefit from advancements in spatiotemporal models are: (i) Self-driving (or autonomous) laboratories [10]. They deploy robots and machine learning-driven sequential decision-making from streaming data to search through high-dimensional parameter spaces (such as process, composition, and property parameters) for materials optimization. A growing number of them are installed at large-scale research facilities such as X-ray or neutron sources or in regular research institutions for organic and inorganic synthesis. (ii) Combined large-scale atomistic simulation and video-mode recording of time-resolved experiments [4]. Here both the simulation and the data analysis may be powered by machine learning algorithms, while data integration between the two modalities through a spatiotemporal model is needed to obtain experimentally validated physical parameters. Both of these two paradigms will benefit from the following developments:

From the model development side, accurately accounting for long-range dependence (LRD) [8] in both spatial and temporal dimensions is one of the crucial yet unmet challenges. LRD manifests in the slowly decaying dependence structure, such that the Markov assumption is no longer a valid approximation. Current approaches using deep-state space models are limited to video frame classification and generation, further improvements on both spatial and temporal LRD, computational efficiency, and the accommodation of graph-structured data will be fitting for the demands in materials data integration.

From the data engineering side, the data integration process often involves the comparison of metadata from two or more sets of measurements or calculations, which require that the data formats are interoperable. Systematic documentation of metadata is crucial for successful data integration projects, which now lie at the center stage of the FAIR principle [1]. For materials optimization platforms that depend on streaming data, the development of automated (meta)data logging systems that include anomaly and distribution shift detection is essential for the quality control of data acquisition. It will also pave the way for efficient data integration and enable online search and process optimization.

**Concluding Remarks**

Spatiotemporal models have demonstrated promising outcomes in integrating data from multiple sources and guiding scientific discovery. The future of spatiotemporal models for materials data science should explore the interplay between the domain knowledge used in problem mapping and model specification to ensure a faithful representation of the problem context to achieve the desired interpretability and performance.

**Acknowledgments**

The work was partially supported by BiGmax, the Max Planck Society's Research Network on Big-Data-Driven Materials Science.

### 3.7: Perspective: Roadmap on Big-Data-Driven Materials Science Soft-Matter Simulations

Tristan Bereau[1] and Kurt Kremer[2]

[1] Institute fur Theoretical Physics, Heidelberg University, Heidelberg, Germany
[2] Max Planck Institute for Polymer Research, Mainz, Germany

**Status**

Soft matter is a sub-class of condensed matter that comprises systems with a characteristic energy on par with thermal energy at room temperature, $k_BT_{room}$ (about $2.5 \cdot 10^{-2}$ eV at T=300K). The low energy gives rise to significant conformational (intra-molecular) flexibility, leading to the spontaneous self-assembly of supramolecular mesoscopic structures. Relevant systems include polymers, colloids, and complex fluids, for which soft-matter physics have provided a foundational understanding [1]. Soft matter offers a slew of modern-day applications, e.g., food products, rubbers for automotives, electronics or medical applications. This makes the discipline both scientifically and technologically highly relevant.

A crucial aspect central is the relevance of multiple scales: phenomena occur at various length- and time-scales, some of which decouple. This simplifies the tackling of complex systems: to build simpler models and focus solely on the relevant degrees of freedom. Scale separation takes its roots in renormalization group theory, and with significant implications in various aspects of theory (e.g., scaling concepts in polymer physics) as well as computer simulations (i.e., multiscale modelling). **Figure 4** illustrates the benefits of multiscale modelling for two applications: high-throughput screening of drug-membrane permeability, and a hierarchical description of polymeric organic electronics.

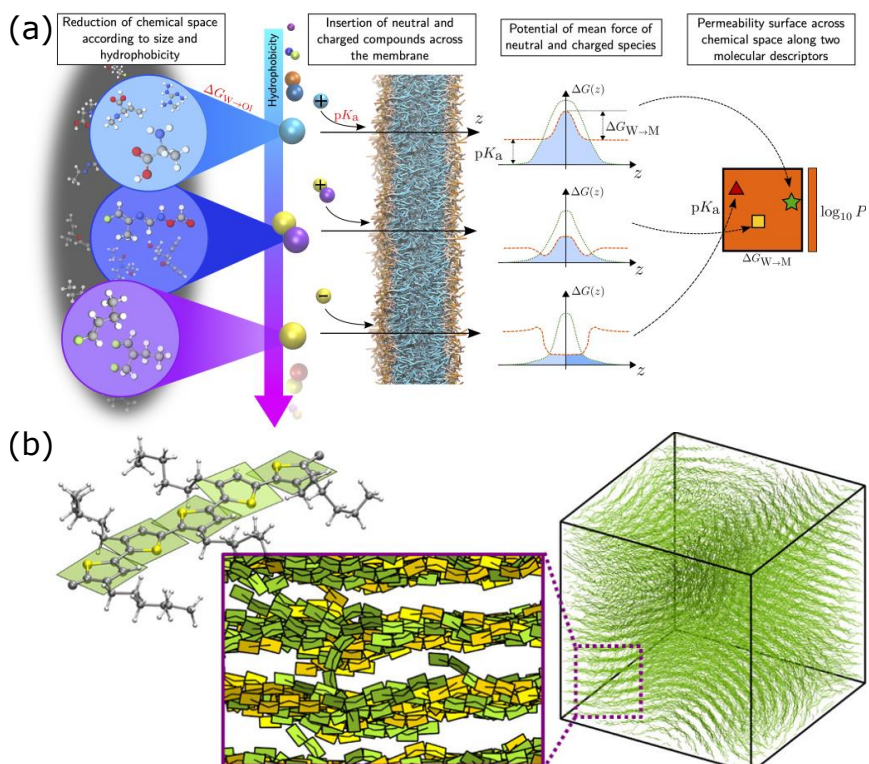

**Figure 4** (a) Coarse-grained simulations of drug-membrane permeability to screen compounds at high throughput. Panel reproduced from [2]. (b) Hierarchy of descriptions of P3HT, a prototypical polymer for organic electronics.

Charges are transported primarily along the backbone of the chains, while the aliphatic side chains are needed to process the material. Panel reproduced from [3].

Soft-matter science has gone through substantial evolution in the last half century. In polymer physics, experiments and theory have worked hand in hand early on to measure coveted critical exponents, and link to general statistical mechanics theory. Computer simulations have played an increasingly important role—they offer invaluable microscopic detail and reach out to ever-growing system sizes [4], [5]. They combine basic generic concepts with specific material properties. In the last decade, data-driven methods, and more recently machine learning (ML), have become increasingly popular in soft matter. They offer an inductive approach to help bridge the scales, and more broadly solve complex structure-property relationships.

Though the penetration of ML in soft matter has been lagging against hard condensed matter, more recent developments show that the outstanding challenges faced by conformational flexibility (i.e., the role of entropy) are increasingly being addressed. In accordance with other fields of physics, chemistry, and materials science, the pursuit of stronger inductive bias (i.e., building physics in the model) systematically helps build better models in an area that is notoriously scarce in data—experimental or from computer simulations. The continued development of ML techniques for soft-matter physics, and the cross-penetration of ML with multiscale modelling, is helping push soft-matter physics toward higher-precision predictive modelling, soft-materials design and optimization, and reproducing entire experiments on the computer [6].

**Current and Future Challenges**

The field of big-data-driven materials science in the context of soft-matter simulations faces several outstanding challenges:

1. The foremost challenge is tackling the "black box" nature of complex machine learning models like deep neural networks. Why does a machine learning model make certain decisions? To this end, interpretability and explainability are paramount. Important developments have been made in the direction of symbolic regression, thereby discovering *mathematical equations* governing the complex phenomena characteristic of soft matter systems. Still, more effort is needed to gather further insight and intuition behind the underlying physics.

2. What makes soft-matter systems fascinating is also what makes them challenging: their multiscale nature. The aggregate effect of many small parts often sums up to large-scale supramolecular behavior—this emergent phenomenon is an outstanding challenge to effectively learn in ML models, and adequately generalize. This is the main reason why computer simulations remain essential nowadays and cannot easily be replaced by ML models alone. Looking to the future, the fusion of ML with physics-based simulation methods (e.g., molecular dynamics or Monte Carlo methods) is expected to have a strong impact.

3. Furthermore, navigating non-equilibrium dynamics stands as a colossal challenge. Almost all soft matter systems—including all of life—exist far from equilibrium. Worse, even systems that appear in equilibrium typically depend on non-equilibrium effects via their *processing*: the mere preparation (e.g., synthesis and subsequent treatment) impacts the final product [7]. The absence of a well-established theory for non-equilibrium statistical mechanics can be an opportunity for inductive methods.

Though, soft-matter science is already traditionally an interdisciplinary field, bringing together physics, chemistry, biology, and materials science, the advent of data-driven methods and ML further reaches out into computer science. The training of scientists that can efficiently work and communicate between these different fields is more important than ever.

**Advances in Science and Technology to Meet Challenges**

Compared to hard condensed matter, soft matter lags behind in terms of ML integration, in large part due to the need to address the associated conformational flexibility. One outstanding challenge lies at the level of system representation, i.e., how to encode the fluctuating system configuration for input to an ML model. Atomic representations developed for electronic properties have focused on single configurations (e.g., [8]). Here instead, observables are averaged over a typically very broad Boltzmann distribution of configurations. Much less work has been proposed in the context of ensemble-averaged ML representations, though ideas have been proposed [9], [10].

Capturing multiscale phenomena lies at the heart of soft-matter physics—from microscopic molecular architecture to mesoscopic structure, to macroscopic behavior. Limitations in the generalizability of ML models strongly limits the current prospects of replacing physics-based models. It is not so clear how extensive the training of an ML model ought to be to reproduce emergent phenomena, such as the self-assembly of soap bubbles from amphiphilic molecules. Coarse-grained modelling has been at the forefront of soft-matter simulations—it exploits scale separation to focus on the most relevant degrees of freedom. Advances in combining coarse-grained modelling with ML is key to further develop data-driven soft-matter simulations. Much work is currently focused on ML-based coarse-grained potentials [11], [12], where striking an adequate balance between accuracy and computational speed is of critical importance. Longer term, it is not clear to what extent ML models might be able to generalize enough to replace the integration of classical equations of motion.

**Concluding Remarks**

It is difficult to overstate the significant impact of first theory, and later computer simulations, on our understanding of soft matter. Bringing soft matter to the fourth paradigm of science (i.e., data-driven methods) will require the tackling of several outstanding challenges. The ongoing developments of machine learning will hopefully continue to naturally evolve from hard condensed matter to soft matter, thereby addressing the needs to model configurational entropy. We foresee that these technical hurdles may help usher soft matter in a new era, where poor scale separation can be efficiently addressed, and insight can be gained for phenomena that are too complex for traditional methods.

**Acknowledgements**

TB and KK thank BiGmax, the Max Planck Society's Research Network on Big-Data-Driven Materials-Science, for support and stimulating interdisciplinary interactions with members of the consortium.

## 3.8: Physics-Enhanced Machine Learning Based Surrogate Modeling for Continuum Mechanics

Pawan Goyal[1, *], Mohammad S. Khorrami[2], Jaber R. Mianroodi[2], Peter Benner[1], Dierk Raabe[2]

[1] Max Planck Institute for Dynamics of Complex Technical Systems, Magdeburg, Germany
[2] Max-Planck-Institut für Eisenforschung, Düsseldorf, Germany
*Email: goyalp@mpi-magdeburg.mpg.de

**Status**

Mathematical modeling plays a pivotal role in the study of continuum mechanics and material design, offering profound insights into material behaviors and microstructures, which in turn, support and guide material optimization and design. Typically, this modeling process involves formulating partial differential equations (PDEs) based on fundamental physical principles such as mass and energy conservation as well as force equilibrium. These PDEs, for given initial conditions and boundary values, are subsequently solved using numerical methods, with finite element and spectral methods being popular choices. Unfortunately, these traditional numerical techniques are computational very costly, a challenge that becomes particularly pronounced when dealing with design studies that require a multitude of simulations under varying configurations. To address this computational burden and streamline the design cycle, there is a pressing need to develop surrogate models that can replace the traditional simulations, often reliant on the methods mentioned above, like finite elements, spectral methods, or finite volume techniques. These surrogate models are particularly valuable during the design phase, offering a more computationally tractable solution.

The use of artificial neural networks (ANNs) in surrogate modeling, driven by advances in machine learning and deep learning, has therefore become a field of growing interest. While neural networks-based material modeling can be traced back to [1], it is in the last decade, with the rapid progress in deep learning and the availability of powerful hardware, that the development of surrogate models using ANNs has surged and continues to expand. Data utilized for constructing these surrogate models can comprise a combination of experimental data, empirical knowledge, and synthetic data generated through numerical solvers. Within the realm of continuum mechanics, numerous methodologies have emerged for building surrogate models using ANNs. For example, in [2] a neural network architecture, namely, conditional generative adversarial networks, has been employed to predict stress and stress fields for a given microstructure geometry; [3] employed a convolutional neural network (CNN) to estimate von Mises stress for microstructures consisting of isotropic elastic and elastoplastic grains, within microstructures, with extensions to heterogeneous periodic microstructures containing inelastic crystal grains in [4] as depicted in Fig. 1. Furthermore, [5] explores the application of the Fourier Neural Operator (FNO) for the surrogate modeling of stress and strain in heterogeneous composites.

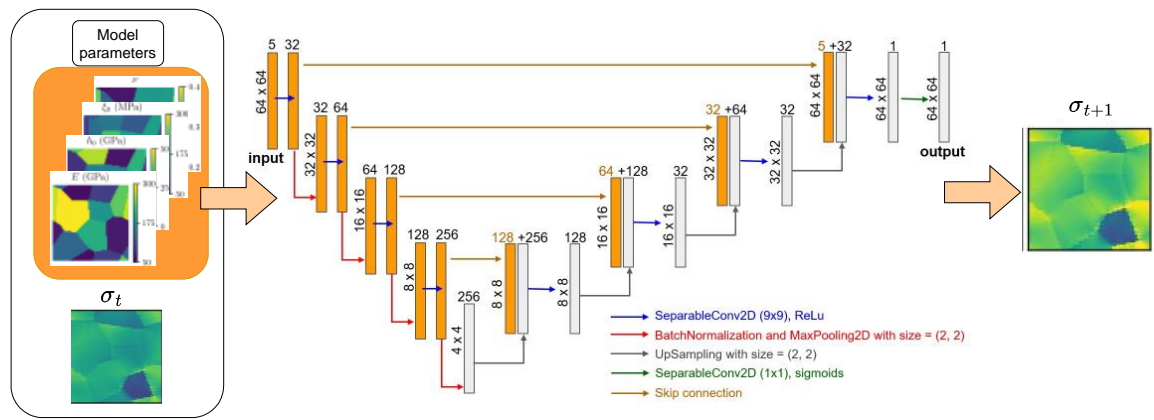

(a) A schematic illustration of the machine-learning-based model.

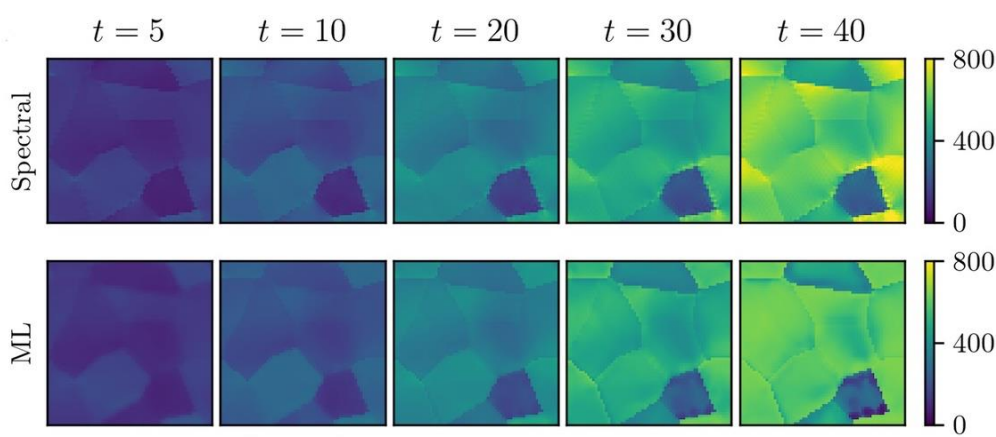

(b) A comparison of prediction quality.

*Figure 5 The schematic illustration of the machine-learning-based surrogate model for predicting the history-dependent local von Mises stress in a solid aggregate that comprises sets of crystalline grains which are characterized here by different elastic-plastic stiffness. The material parameters considered (varied from grain-to-grain) in the simulations include E: Young's modulus, $\mu$: Poisson ratio, $\zeta_0$: initial inelastic flow resistance (viz. plastic deformation), $h_0$: initial isotropic hardening, and $\sigma$: von Mises equivalent stress. We construct a surrogate model of stress fields for visco-plastic polycrystalline materials using the U-Net architecture as shown in (a), which predicts von Mises stress field 500 times faster than conventional spectral solvers, see (b). The figure is modified from [3] and [4] with permission.*

**Current and Future Challenges in Using Big Data Methods for Continuum Mechanics**

Often, surrogate modeling is conducted purely based on large amounts of data, mostly by training ANNs with them. However, within the context of continuum (micro-)mechanics, there exists a wealth of established physical and empirical knowledge [11] that ANN-based surrogate methods have yet to fully incorporate. In the following, we discuss the notable challenges in bridging this gap for surrogate modeling in continuum mechanics.

a) *Physics-enhanced surrogate modeling*: Incorporating physics-based knowledge into surrogate modeling is an active research field. For instance, in [6] and [7], physics-based knowledge, including the underlying PDEs and empirical knowledge, has been leveraged to introduce biases into ANNs, resulting in outputs that approximate the underlying physics, such as enforcing divergence-free conditions as well as mass and energy conservation. However, it is essential to note that these approaches primarily aim to satisfy the physical laws in a weak sense. Hence, the output from the trained ANNs may not be fully physically meaningful, particularly at a local scale. Therefore, we need to explore the design of neural network architectures that are capable to inherently produce an output that satisfies physics in a strong sense, with a particular focus on critical properties like divergence-free behavior, as well as mass and energy conservation both, on a global and local scale.

b) *Stable dynamic prediction*: Surrogate modeling has been used for predicting time-dependent stress and stress fields of heterogeneous solids subject to homogenous steady-state external loading conditions. Within this framework, these surrogate models can be regarded as dynamical systems. Given that surrogate models typically emulate stable physical behavior thereby mimicking the basic rules of continuum mechanics, it is essential for them to possess inherent stability, i.e. mimicking also convergence. This stability ensures that predictions remain consistently stable and bounded. Consequently, it is imperative that ANN-based surrogate models are designed to have these stability properties inherently embedded.

c) *Learning low-dimensional latent representation*: Often, the field of interest in continuum mechanics is two or three-dimensional real space, ideally also informed by the solid's crystal and phase state, adding further dimensions and anisotropy features to the problem to be solved. Consequently, the data obtained for these scenarios are high-dimensional, especially while dealing with high-resolution spatial fields. However, it is a common observation that such high-dimensional solutions can often be accurately represented in a low-dimensional latent space. The creation of this low-dimensional space is further guided by constraints designed to simplify the dynamics and engineering design processes. For instance, it is possible to construct a latent space in such a way that the system dynamics evolve in a nearly linear fashion, aligning with principles like Koopman theory and dimensionality reduction techniques.

**Advances to Meet these Challenges**

In our pursuit of designing neural network architectures that inherently adhere to physical properties (e.g., divergence-free, energy preserving), we seek to utilize fundamental mathematical vector calculus. For illustration, in order to design ANNs to produce divergence-free quantities, we seek to obtain intermediate quantities so that divergence-free are obtained by taking the curl of those intermediate quantities. Such techniques find widespread use in solving PDEs (e.g., Maxwell equation) with divergence-based constraints. Additionally, for achieving stable time evolutions through neural networks, we extend concepts proposed in [8] to encompass high-dimensional spatial and temporal data. Furthermore, our empirical studies indicate that CNNs that explore local features underperform compared to FNO, which explore global features present in the data. Therefore, our exploration centers on incorporating these physical properties within the context of FNO. We further need to explore how these trained networks are used for engineering studies such as predicting optimal material property configurations, drawing inspiration from [9]. What is more, we seek to discover suitable low-dimensional latent representation through autoencoders with the intent to simplify the task of predictions and engineering studies. Algorithmic developments in this direction have been pursued in [10], which requires further investigation in the context of continuum mechanics.

**Concluding Remarks**

We conclude by emphasising that it is imperative to develop new machine learning and deep learning methodologies for tackling problems pertaining to the continuum mechanics of heterogeneous and anisotropic solids that adhere to the strong forms of essential physical principles both on a global and on a local scale. Doing so offers several advantages: firstly, it enhances the interpretability and generalizability of machine learning-based surrogate models. Secondly, it reduces the amount of required training data. Thirdly, it can enhance solver performance by up to several thousand times compared to conventional solution methods such as FEM or spectral methods. As an initial endeavour in this direction, we have demonstrated how to construct machine learning surrogate models that inherently produce divergence-free stress fields, thereby satisfying mechanical equilibrium conditions. Learning suitable low-dimensional latent representations not only reduces online inference time but also facilitates engineering studies with minimal computational resources. Additionally, acquiring training data for engineering applications is both economically expensive and time-consuming. Therefore, it is crucial to devise strategies for cleverly gathering training data, ensuring that the limited data covers a wide range of parameter space.

**Acknowledgements**

The authors acknowledge funding by BiGmax, the Max Planck Society's Research Network on Big-Data-Driven Materials Science.

## 3.9: Digitalization of Advanced Experimental Techniques for Microstructures

Christoph Freysoldt[1], Baptiste Gault[1,2], Christian H. Liebscher[1], Pawan Goyal[3] and Jörg Neugebauer[1]

[1] Max-Planck-Institut für Eisenforschung GmbH, Düsseldorf, Germany
[2] Department of Materials, Imperial College, London, UK
[3] Max-Planck-Institut für Dynamik komplexer technischer Systeme, Magdeburg, Germany

**Status**

Most modern engineering materials exhibit a complex microstructure that underpins the properties of the material in beneficial – or sometimes detrimental – ways. This applies to structural alloys, to ceramic materials like concrete or protective coatings, as well as to functional materials for energy storage, electronics, heterogeneous catalysis, etc. Steels, for example, consist of several meta-stable phases formed during casting, thermo-mechanical processing, or in operation. To image the interplay of grain morphology and texture, chemical composition, crystallographic relationships, and local properties of distinct regions, various complementary 'imaging' experiments are available, such as electron microscopy, atom probe tomography, beam diffraction (electron, X-ray, synchrotron), or spatially resolved spectroscopy. Thanks to progress in experimentation, data storage capacities, and digital data processing, these techniques yield an ever-increasing data pool. A single experiment can provide GBs or even TBs of data, which is further multiplied by high-throughput experimentation or *in situ* monitoring of transformations that add a time dimension. This big data is both a challenge and a great opportunity for data-driven research.

So far, it is mostly up to human experts to identify the microstructural features of interest within the experimental data. Often, it is not clear *a priori* what features relate to performance in the applied context. Once identified, one would like to quantify their number density, size distribution, chemical characteristics, and functional properties, in order to extract quantitative processing-microstructure-property relationships that facilitate material design. To automate this process, pattern recognition algorithms are actively being developed [1, 2, 3], often specifically targeted at a particular experiment for a particular type of material. Upon success, they provide a secondary characterization of the material in a reproducible and scalable way. This becomes particularly attractive when combined with high-throughput experimentation to systematically explore a material space.

Merging such derived data, possibly even from different experiments, with traditional materials' characterization across multiple samples while tracking their synthesis and processing history alongside necessitates a careful data management. Electronic lab books [4], integrated work-flow environments [5], structured material databases [6], and flexible data sharing platforms [7] cover some, but not all aspects. The barriers between them effectively limit data-driven material's design.

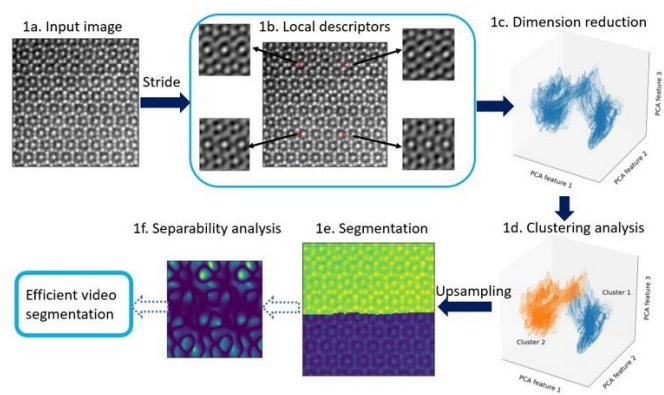

**Figure 1** Crystallographic segmentation of atomically resolved STEM-HAADF frames via symmetry descriptors, clustering, and distinguishing feature selection for enhanced performance (taken from [1] with permission).

**Current and Future Challenges**

Suitable algorithms for pattern recognition are available from other fields, but must be adapted to a specific research question, see Figure 1. Exploiting domain knowledge to define suitable descriptors and selecting robust algorithms [1,2] will remain a scientific challenge in the coming years due to the vast variety of relevant phenomena and patterns. The actual integration of automatic microstructure evaluation in research practice is still at infancy. Progress is presently hindered by:

**(1) a lack of established data and file formats.** Experimental raw data is typically acquired in instrument-specific file formats. Extracting all potentially relevant data for machine-learning workflows is often not possible or impaired. Community efforts have been undertaken to establish open data formats [8,9]. An alternative effort aims at read-function libraries that support multiple formats [10]. For storing analysis output, ideally in conjunction with input data, no standard exists. Similarly, exchanging data between different data management systems is severely hindered by inherent heterogeneity in data structures and metadata, in the naming and unit convention of data fields, and by assuming implicit context (e.g., providing an instrument's name rather than its measurement parameters).

**(2) a lack of flexible workflows or tool chains.** Material science research routinely combines different characterization methods, but rarely so in a digitally integrated way. Researchers have their individual ways to document a material's synthesis and processing history, how each experiment's specimen was prepared, and how data was post-processed. Common approaches (via file name, free-form notes, folders, …) are ill-suited for automatic processing. Electronic lab-book systems help to manage those data [4], but typically reach their limits in collaborations across labs.

**(3) a cultural gap between experimentalists accustomed to graphical user interfaces (GUIs) and programming-oriented data scientists.** Present-day analysis strongly relies on humans to inspect the data. Instrument manufacturers therefore provide monolithic visualization tools with a GUI, that read the instrument's raw files, provide a fixed set of processing schemes and export results in established general-purpose image (jpg, tiff) or data formats (csv, hdf5) that drop context. In contrast, the wider machine-learning field thrives on plugging together open-source libraries and code snippets on demand, that require significant coding skills.

**Advances in Science and Technology to Meet Challenges**

To reconcile the cultural gap, today's interactive data visualization and future advanced data processing must be interlinked. GUI-based visualization tools could open up by establishing plug-in mechanisms to exchange data and visualization items with external modules. An alternative route, that circumvents the GUI integration challenges, is to follow the successful model in *computational* material science [5, 7]: focus on input/output data format normalization, and employ separate tools that work with these formats for analysis and visualization, all coupled together by a managing framework, see Figure 2. Further efforts to standardize input, but more importantly for recurring output such as classification signatures, segmentation maps, interface location, geometric shape information, etc. are urgently needed.

In this context, exploiting automatic code generation from machine-readable data format definitions - in conjunction with ontologies and knowledge graphs - could be a game-changer to speed up the development, as they reduce the human effort in defining standards *and* implementing corresponding code for possibly different programming languages. Similarly, the trend towards higher abstraction in machine-learning software should be exploited to generate processing metadata. When the transformation chain is built at run-time via high-level objects (which later generate the actual code for the hardware at hand on the fly), the high-level representation should automatically annotate the data output with the details of the processing chain.

At a higher level, workflow and data management tools must be adapted to deal with the specific challenges of experimental data. As experimental data sets can become very large, moving or copying around entire data sets is prohibitive. In most cases, raw data will be stored close to where it was generated. Computational resources for advanced machine learning might be located elsewhere, and only need part of the data, or specifically pre-processed data that reduces transfer size via dimensionality reduction or compressed sensing. Thus, workflows that deal with both distributed data and distributed computation will be needed, while maintaining consistency in metadata and ensuring that data access across computer systems is reliably authenticated to avoid premature publication or leakage of confidential data.

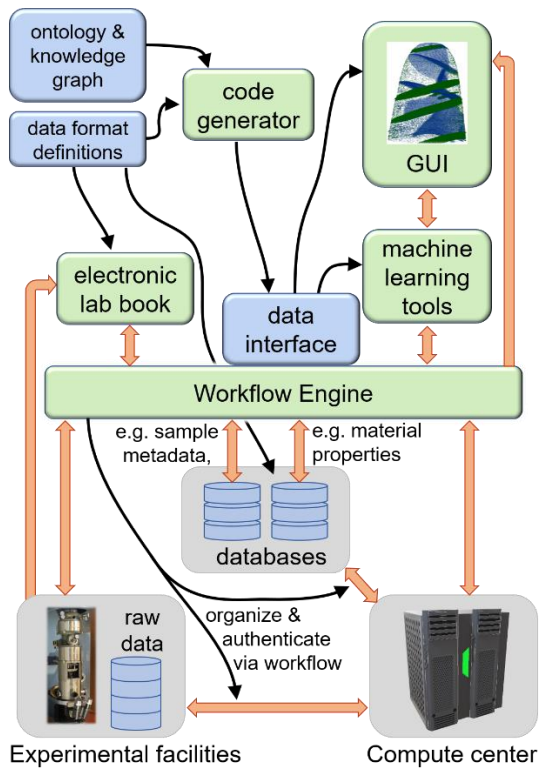

**Figure 2** Sketch of a possible digital infrastructure for handling data from experimental imaging techniques in an integrated workflow.

**Concluding Remarks**

The success of data-rich imaging techniques in material science lies in the promise that materials' properties are linked to recurring patterns that can be discovered by inspecting a few representative examples. Machine-learning techniques can leverage this approach by removing the human inspection as the limiting factor to digest larger and larger amounts of data in order to discover relevant, but possibly rare patterns. At the same time, they offer the unique chance to characterize the underlying distributions in a statistically significant manner as more data becomes available, thus generating secondary high-level characterizing data that might serve as valuable descriptors for associated properties. Digitalizing the entire workflow from synthesis, sample preparation, data acquisition and post-processing in an integrated way as sketched in Fig. 2 is critical to achieve these goals.

**Acknowledgements**

This project received financial support from BiGmax, the Max Planck Society's Research Network on Big-Data-Driven Materials Science.

## 3.10: Opportunities and Pitfalls of Using Large Language Models in Materials Science and Engineering

Dierk Raabe[1], Zongrui Pei[2], Junqi Yin[3], James Saal[4], Kasturi Narasimha Sasidhar[5], Jörg Neugebauer[1]

[1] Max-Planck-Institut für Eisenforschung GmbH, Düsseldorf, Germany
[2] New York University, New York, USA
[3] Oak Ridge National Laboratory, Oak Ridge, TN, USA
[4] Citrine Informatics, Inc., Redwood City, CA, USA
[5] University of Wisconsin, Madison, USA
* Email: d.raabe@mpie.de

**Status**

Large Language Models (LLMs) could grow into a transformative research tool in materials science, bridging the gap between text-based data and actionable insights. Opportunities lie in accelerated materials synthesis, discovery, processing and property design. LLMs can be used as indirect or direct tools. By indirect, we refer to situations where LLMs help in extracting data and building databases from scattered sources. This does not involve the actual process of materials discovery but it is a precursor step. Such databases can then be used by other machine learning (ML) methods. With their direct role in materials discovery we mean that LLMs can even extract causal relationships from collected data, serve to build domain-specific knowledge graphs, render hypotheses and guide progress-critical experiments, data collection and simulation (1–3). The latter aspect is essential because LLMs do not obey any built-in causal rules. Instead, they connect language tokens in a probabilistic way, without considering logic, self-consistency or conservation laws. This means that they can violate elementary scientific rules. They mimic scientific context by using probability measures that rest on majority but not on proof or logic. This explains why there are opportunities but also pitfalls. The latter can be mitigated by combining LLMs with other methods such as classical theory, thermodynamics, kinetics, materials property data bases, explainable artificial intelligence, active learning etc. LLMs are also capable of generating hypotheses and they can be used to build domain-specific knowledge graphs which in turn can enhance predictive models (4, 5).

Materials science stands at the confluence of several disciplines. Research topics range from latest quantum mechanical insights into the behavior of electrons in complex systems to large-scale processing of billions of tons of material (concrete, steel) and materials exposed to harsh environmental conditions (catalysts, corroding products). Developing data-centric methods to leverage disruptive progress in this field must, therefore, reflect and embrace this heterogeneity in the underlying data from which knowledge can be extracted, combined and used.

In the portfolio of model-based artificial intelligence (AI) methods, LLMs seem to offer new opportunities to discover materials and processes that may otherwise remain hidden in the complexity and scattered information that already exists (6). One avenue to use LLMs is accelerated materials discovery (7–9). This is due to the fact that language-based token systems that connect words based on probability are particularly strong in extracting and combining knowledge that already exists in text form. Therefore, while LLMs may not be necessarily suited for disruptive conceptual discoveries from text connections, they can accelerate design based on existing concepts (10). Although this is a rather conservative approach, it is already a big step forward, because the traditional trial-and-error approach of material discovery is time-consuming and resource-intensive. Also, LLMs can analyse vast datasets, extracting patterns and correlations that would elude human researchers. For instance, LLMs can process published literature, patents, and experimental data to suggest combinations of novel material compositions and even possible properties, as will be shown below in more detail. By integrating databases like the Materials Project or the Cambridge Structural Database, LLMs can offer quantitative predictions about material structures, compositions, and potential applications, significantly reducing the time from conception to application.

However, it should also be noted that Krenn and Zeilinger (11) recently suggested a more disruptive approach to use LLMs. They introduced SemNet, a dynamic knowledge organization method in the form of a continuously evolving network, constructed from 750,000 scientific papers dating back to 1919. Each node in SemNet represents a physical concept, and a link is established between two nodes when the concepts are jointly explored in articles. SemNet has proven its utility by enabling the authors to pinpoint influential research topics from the past. The authors trained SemNet to forecast trends in quantum physics, and these predictions have been validated using historical data.

A few examples of using LLMs in materials science have been recently presented. Jablonka et al. (2) conducted a hackathon using LLMs such as GPT-4 for chemistry and materials science. The participants leveraged LLMs for a variety of purposes, such as predicting properties of molecules and materials, creating new tool interfaces, extracting knowledge from unstructured data, and developing educational applications. Being more specific, An et al. (12) argued that the construction of knowledge graphs for domain-specific applications like metal-organic frameworks (MOFs) can be resource-intensive. LLMs, particularly domain-specific pre-trained models, have been successfully employed to create such graphs. For example, a study explored the use of state-of-the-art pre-trained general-purpose and domain-specific language models to extract knowledge triples for MOFs (12). The authors constructed a knowledge graph benchmark with 7 relations for 1248 published MOF synonyms. Experimental probing revealed that such domain-specific pre-trained language models (PLMs) outperformed general-purpose PLMs for predicting MOF related triples. The authors also conceded from their overall benchmarking results that the use of PLMs alone to create domain-specific knowledge graphs is still far from being practical and requires the development of better-informed PLMs for specific materials design tasks. The group of Olivetti used LLMs to generate knowledge graphs (MatKG2) for the entire domain of materials science, taking ontological information into account as opposed to using statistical co-occurrence alone (14). Zhao et al. (13) used fine-tuned Bidirectional Encoder Representations from a Transformer (BERT) model and tested it with respect to data extraction from published corpora. They reported that the model achieved an impressive F-score of 85% for the task of materials named entity recognition. The F-score is a metric used to evaluate the accuracy of a model in binary classification tasks. Sasidhar et al. (16) integrated natural language processing and deep learning for the design of corrosion-resistant alloys (17). They also highlighted the general challenges in utilizing textual data in machine learning models for material datasets and proposed an automated approach to transform language data into a format suitable for subsequent deep neural network processing. This method significantly improved the accuracy of pitting potential predictions for alloys, providing insights into the critical descriptors for alloy resistance, like configurational entropy and atomic packing efficiency. Pei et al. (10) proposed a concept of 'context similarity' to select chemical elements with high mutual solubility for discovering high-entropy alloys. They trained a word-embedding language model with the abstracts of 6.4 million papers to calculate the 'context similarity'. With this approach they designed a workflow to design lightweight high-entropy alloys, which suggested even 6- and 7-component lightweight high-entropy alloys by finding nearly 500 promising alloys out of 2.6 million candidates.

Gupta et al. (8) developed MatSciBERT, a materials domain-specific language model for text mining and information extraction. They argued that conventional language processing alone, such as encoded in the form of BERT models, may not yield optimal results when applied to materials science due to their lack of training in materials-specific notations and terminology. To address this challenge, the authors introduced a specific materials-aware language model they refer to as MatSciBERT. This model was trained on an extensive corpus of peer-reviewed materials science publications. The authors claimed that their model surpasses SciBERT, a large language model trained on a broader and less materials-specific scientific corpus, in three critical tasks, named entity recognition, relation classification, and abstract classification. The developers made trained weights of MatSciBERT publicly accessible, enabling accelerated materials discovery and information extraction from materials science texts. A recent study introduced a larger GPT version, named MatGPT (18), based on a larger scientific corpus than MatSciBERT. In their study the group claim that the MatGPT model embeddings outperform MatSciBERT and achieve an improved band gap prediction based on the Materials Project combined with graph neural networks (GNN).

**Current and Future Challenges: LLMs and Knowledge Graphs for Materials Discovery**

The current flagship in the world of LLMs is the Generative Pre-trained Transformer 4 model (GPT-4) from OpenAI. It is based on 8 separate models, each containing dozens of network layers and 220 billion parameters, which are supposedly linked together using the Mixture of Experts (MoE) architecture. GPT-4 is built on a transformer architecture, combining self-attention and feed-forward neural networks to process input tokens. Each token represents a text string containing a word or phrase. Therefore, the token limit represents the amount of text that an LLM can consider at a given time as input. Early LLM releases had very low token limits since LLM calculation time is strongly dependent on the token length. Initial releases of GPT3 had a token limit of 2,048 tokens, but recent releases of GPT-4 has a token limit of 128,000 tokens. To put this into context, the average length of a PubMed abstract is 114 tokens (sd 48.83) and an article is 2,378 tokens (sd 1,604.79). So while increasing complexity of the LLMs has enabled using entire papers (or even groups of papers) as input, there is still the computational cost of running GPT-4 calculations to consider. As an example, when asking a question of medium complexity via a string of fewer than 10 tokens, such as 'Composition and property ranges of material XY', then the rough total cost estimate to answer this question for GPT-4 is about 7-10 Euros. Getting the same answer from a classical knowledge graph would incur only about one-hundredth of this cost and also take less time, provided the information is in the corpora and mapped in a graph accessible by search engines.

Using knowledge graphs also removes the hallucination effect, an error made by LLMs when rendering combinations that appear plausible to the model's probability measures but false when tested against high-fidelity information or logic. It appears due to multiple factors, such as when LLMs are trained on contradictory datasets, overfitting, etc. An urgent and vital topic in LLMs is, therefore, quantifying the level of the hallucination effects and developing a systematic method to recognise and mitigate them. On the other hand, LLMs have the advantage that they can process and understand the context from scientific literature, patents, and database entries. When combined with knowledge graphs that structure this information, it provides a rich database of materials science knowledge which can be readily queried. This integration allows for the rapid assimilation of existing knowledge and the identification of knowledge gaps.

Vice versa, LLMs, with their ability to process and generate large volumes of text, can also serve to construct domain-specific knowledge graphs, optimize algorithms for faster discovery, and enable more efficient design and exploration of materials. The synergy between LLMs and knowledge graphs could hence be a useful next step to materials discovery, offering a paradigm shift from traditional, iterative experimental methods to a more quality-controlled data-driven model. This combination would allow better alignment of reliable high-quality data exploitation (through knowledge graphs) and semantic contextualization (through LLMs).

LLMs can also analyse patterns and relationships within a knowledge graph to generate hypotheses-based suggestions for suitable search spaces pertaining to potentially novel materials and properties. For instance, by understanding the relationship between crystal structure and electronic properties, LLMs coupled to knowledge graphs could likely be used to suggest new compositions or corresponding search spaces for magnets, battery materials or solar cell absorbers.

**Advances to Meet Challenges associated with the use of LLMs in Materials Science**

While the opportunities are vast, applying LLMs in materials science also has challenges. Data quality and availability are critical as models are only as good as the data they train on. Ensuring data integrity and representativeness is paramount. Furthermore, the interpretability of LLM outputs is crucial for gaining trust in their predictions. Developing models that can provide not just predictions but also insights into the underlying mechanisms is an essential goal. Another point is the Chain of Thought Prompting, an approach to enhance LLMs' comprehension of causal relationships and reduce hallucination. It involves forcing the models to verbalize different steps of reasoning they have gone through in reaching conclusions. This makes the process more transparent. Such ideas have not been implemented in materials science but in other areas such as medical science (1).

The quality of the information that can be extracted from LLMs depends on the quality and timeliness of the input text. For material science that can be only achieved if the latest literature that has been going through proper peer review processes is being used. However, only one-third of the current scientific corpora is open access. Therefore, some of the corpora currently used for training LLMs is in part of questionable quality. Also, current LLMs might simply miss the latest literature. This means that the model weights are not fitted to the latest state of the art. These two aspects show that fine-tuning prior to the use of such LLMs is recommendable. On the other hand, recent literature sometimes also overlooks knowledge that already exists long in the literature so that some findings reported in papers are more like re-discoveries, a problem that can be likely mitigated when LLMs are used. In this context, is it worth that Application Programming Interfaces (APIs) being now offered by a few companies to allow accessing millions of publications along with metadata. Another issue is that extracting text from PDF files, the standard format of the literature, results in poorly formatted corpora with numerous errors (e.g., missing text, insertion of text from other items such as tables in the middle of sentences, headers and page numbers, etc.).

An unresolved open front of LLMs is the potential violation of existing copyright when tapping into web-based resources, which becomes an obvious issue with the use of journals, textbooks, and other scientific literature in training. Another concern is if further tuning of LLMs leads to slow asymptotic knowledge increase because high-quality peer-reviewed content on certain topics is not growing at a sufficiently high rate and is often not freely accessible for training. In other words, it is not likely that LLMs can gain knowledge quicker than the generic basic research used to train them. To meet both challenges, the rapidly growing fraction of open-access literature and the use of pre-publication and self-archiving services is of great value, likely leading to higher quality improvement and less hallucination of LLMs. Some of these aspects also connect to general limit considerations regarding model capacity and scaling laws, which were recently shown to depend essentially the number of model parameters, the size of the dataset and the amount of computation power used for training. Performance was shown to depend less on other architectural hyperparameters such as depth and width. However, irrespective of these theoretical considerations, the scientific community has not yet seen the capacity limits of the GPT model in current applications. This means that for the same data size, the GPT model improved further as the number of parameters was further increased.

**Concluding Remarks**

LLMs offer great potential in the complex interplay between advanced computational methods and the nuanced, often experimentally and empirically grounded field of materials science. Opportunities lie in accelerated material discovery; enhancement and improved pattern and result analysis of data obtained from existing computational tools such as atomistic simulations; better knowledge synthesis and data management from research articles, reports, and property studies; support in hypothesis development and outlier analysis; and advanced decision-making support in materials selection and design, including aspects such as costs, sustainability and regulatory constraints. Pitfalls exist regarding the quality, availability, bias and legal status of the training data; lack of built-in logic or conservation laws; lack of the reflection of microstructure, synthesis, sustainability and processing complexity; and the danger of over-reliance and even complacency regarding LLM predictions, i.e. the decay of individuals' own critical thinking, rigorous validation or falsification and the thrive towards deep understanding of the underlying causality behind phenomena which are key factors that have made the scientific method the most successful and reliable approach in history.

This contemplation about a few generic pro and con aspects shows that while LLMs offer transformative potential in materials science, their successful integration into the field necessitates careful consideration of the quality and completeness of the data they are trained on, a thorough understanding of the underlying physical and chemical principles, and a balanced approach to leveraging their computational power with critical human expertise.

**Acknowledgements**
The authors acknowledge funding from BiGmax, the Max Planck Society's Research Network on Big-Data-Driven Materials Science.

# Section 4: Material Discovery and Applications

## 4.1: High-Throughput Materials Discovery with AI-Guided Workflows

Thomas A. R. Purcell[1,2], Luca M. Ghiringhelli[1,3], Christian Carbogno[1] and Matthias Scheffler[1]

[1] The NOMAD Laboratory at the FHI of the Max-Planck-Gesellschaft and IRIS-Adlershof of the Humboldt-Universität zu Berlin, Germany
[2] University of Arizona, Biochemistry Department, Arizona, USA
[3] Department of Materials Science and Engineering, Friedrich-Alexander Universität, Erlangen-Nürnberg, Germany

**Status**

Computational, high-throughput materials discovery is seen as a promising route to advance a myriad of technologies including batteries [1], renewable energy [2], and pharmaceuticals [3]. With the increasing amount of computer power over the past several decades, millions of materials' properties were calculated on hundreds of thousands of materials, with the aid of high-throughput workflow. Such workflows allow a user to define a set of calculation parameters and run those calculations for a large set of materials. The results then populated several large databases, e.g. Materials Project, AFLOW, Open Quantum Materials Database, NOMAD, etc. [4]. However, as the materials space is practically infinite, such studies can only address a marginal part of it, even for relatively simple properties.

The computational funnel model [5] extends high-throughput studies to complex materials properties by screening out materials after each step according to selection criteria based on the expected result. In theory, this means that the costliest calculations or experiments are done only for the most promising candidates. Naturally, this process is the more successful, the faster and the more reliable undesired materials can be disregarded.

Active learning provides one way of combining AI models and high-throughput workflows [6] The goal of this algorithm is to balance exploitation and exploration to select new data points that can optimize a property or better train an AI model for a given material property, while simultaneously finding a global optimum for it in a data-efficient manner. By using an acquisition function that balances exploration and exploitation to select which materials to calculate next, these frameworks can improve the efficiency of the studies by selecting which materials enter the funnel in a non-subjective manner (see the roadmap paper by Boley et al.). In essence, both the active learning framework and the selection funnels attempt to achieve the same goal, that can be achieved synergistically: Active learning suggests which materials enter the workflows and the high-throughput funnel removes the unpromising candidates after each step of the workflow.

**Current and Future Challenges**

The main challenge in fully realising the potential of AI-guided workflows is integrating active learning schemes, and the AI models and the suited acquisition function they are based on, into advanced selection funnels. The criteria used for each step of the funnels are either based on an expected error bound of a lower accuracy calculation or a physics-informed heuristic, e.g., a material having too large of an electronic band gap or being too dense. The end goal of the screening criteria is to reduce the overall cost and time of a study, while still exploring over the relevant parts of materials space.

While useful, the current screening criteria are a potential obstacle when combining active learning with high-throughput workflows. Because they are not necessarily derived from the data that underlies the AI-model, an overzealous screening procedure can exclude materials that would drastically improve model performance and possibly correct an initial bias. Importantly, the heuristics used to screen out materials may not directly relate to the target property, but be controlled by an unknown third process, leading to an incorrect physical interpretation. Furthermore, adding selection funnels to active learning frameworks could perpetuate the initial bias of the models as the dataset will be directed towards the existing conditions. One potential solution to this problem is through using multi-objective learning to simultaneously optimize both the screening criteria and the target property. However, a less complex solution would be preferable.

The final challenge with creating these workflows is to incorporate them into existing materials discovery frameworks. Currently, the tools used for materials discovery such as AFLOW [7], atomate [8], and Aiida [9]. Without native integration, multiple, potentially incompatible solutions must be created leading to a less transparent ecosystem. An additional benefit of fully integrating these methods is an improved selection procedure. The use of cost-aware and efficient acquisition functions is becoming increasingly popular, [10] and including the AI model training and selection steps inside the workflow libraries themselves will improve the estimated costs for these acquisition functions and multi-fidelity approaches.

**Advances in Science and Technology to Meet Challenges**

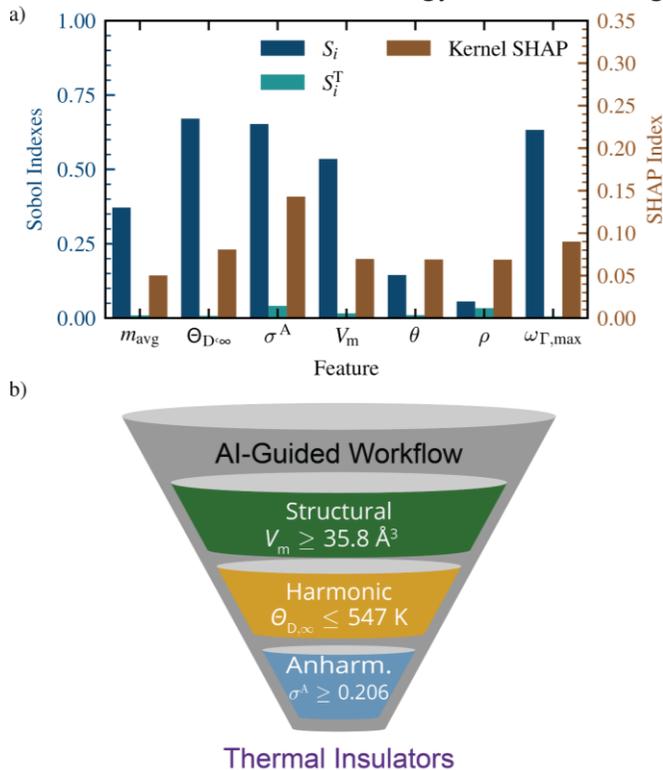

**Figure 1** Example of the proposed workflows. a) The SHAP (brown bars) and Sobol indexes (blue bars) for the AI model found in [Purcell2023] for a model of the thermal conductivity of a material using its structural (average mass, $m_{avg}$; density, $\rho$; molar volume, $V_m$; and reduced mass, $\mu$), harmonic properties (Debye temperature, $\Theta_{D,\infty}$, and the maximum $\Gamma$-point frequency, $\omega_{\Gamma,max}$), and the anharmonicity factor, $\sigma^A$, b) The workflow obtained from the expected thermal conductivity for a given input features of the most important inputs. Adapted from Ref [11]

An expanded use of explainable AI methods presents a clear path to achieve the necessary combination of methods presented above. By learning the conditions for screening out materials from the AI models themselves, explainable AI attempts to expands the predictive power of machine learning models, and give insights into the relationship between the input physico-chemical materials features and target properties. These methods can relate to either the regression method used, e.g. linear or symbolic regression, or post-processing techniques that uncover the relationships. By better understanding the connections between the input features and a target property, one can then replace the physics-derived heuristics with ones from the model itself.

We recently demonstrated the capabilities of this approach, by creating an AI-guided workflow for finding thermal insulators [11]. For this project we modelled the thermal conductivity, $\kappa_l$, of a material based on its structural, harmonic, and the anharmonic properties (see [11] for a complete list). We then applied feature importance metrics, and found only three inputs were important. From here we were able to map the expected value of $\kappa_l$ against each of these inputs to find the screening procedure highlighted in Figure 1b. With this workflow we were able to efficiently find 16 predicted ultra-thermal insulators with a $\kappa_l$ less than 1 W/mK out of an initial set of 732 materials [11].

To fully address the challenges associated with creating sustainable, AI-guided workflows, active learning techniques must be integrated into them. While the selection funnels can find a list of hundreds of possible candidate materials, it cannot identify which predictions are the most important to validate next. However, introducing an acquisition function the workflows can then maximize the quality of information gained per calculation or experiment. In turn this will allow us to speed up the discovery of good materials for vital applications. More importantly, by redoing the feature importance study after each iteration we can further refine the screening criteria and continue calculations that were initially discarded because they broke one of the old metrics.

**Concluding Remarks**

AI-guided workflows have the potential to revolutionise materials discovery frameworks by focusing calculations or experiments on the most promising materials, and potentially remove the initial bias of data selection. By using an appropriate acquisition function to determine which experiments or computations to run next, we can automate these calculations. In turn the focus of the researchers working on these problems can instead be on further developing new methods and not managing a large set of calculations. Furthermore, explainable-AI methods will help elucidate why the models are deciding which candidates to calculate next. With this insight, part of the physical mechanisms driving, facilitating, or hindering the different processes may also be understood. Finally, as the frameworks become better focused the overall efficiency of these efforts will be significantly enhanced.

**Acknowledgements**

This project received financial support from BiGmax, the Max Planck Society's Research Network on Big-Data-Driven Materials, the NOMAD Center of Excellence (European Union's Horizon 2020 research and innovation program, Grant Agreement No. 951786) and the ERC Advanced Grant TEC1p (European Research Council, Grant Agreement No. 740233).

## 4.2: Roadmap for Big Data and Artificial Intelligence Driven Data Analytics in Scanning / Transmission Electron Microscopy (S/TEM)

C.H. Liebscher[1], G. Dehm[1], C. Freysoldt[1], A. Leitherer[2,3], L.M. Ghiringhelli[2,4]

[1] Max-Planck-Institut für Eisenforschung, Düsseldorf, Germany
[2] The NOMAD Laboratory at the Fritz-Haber-Institut of the Max-Planck-Gesellschaft and IRIS-Adlershof of the Humboldt-Universität zu Berlin, Berlin, Germany
[3] Present address: ICFO-Institut de Ciencies Fotoniques, The Barcelona Institute of Science and Technology, Castelldefels (Barcelona), Spain
[4] Department of Materials Science and Engineering, Friedrich-Alexander Universität, Erlangen-Nürnberg, Germany

**Status**

Recent developments in aberration-corrected electron optics, spectrometer and detector technologies enable to capture multimodal signals within a single experiment in a scanning / transmission electron microscope (S/TEM) down to the atomic level. These advancements have greatly expanded our understanding of the atomic constitution of materials, which is largely driven by the rich and multimodal data streams. Spectroscopic techniques such as energy dispersive X-ray (EDS) or electron energy-loss spectroscopy (EELS) can nowadays probe the local composition and electronic structure of complex materials at the atomic level. New scanning diffraction methods, termed 4D-STEM, capture 2D electron diffraction patterns in each probe position of the 2D raster grid and have facilitated to image light elements at atomic resolution, determine local structures and strain with sub-nanometer precision [1]. Further, the spectroscopic and 4D-STEM techniques can be combined with tomographic approaches to obtain the 3D nature of materials. Advances in *in situ* probing capabilities and fast electron detectors make it possible to directly observe the dynamic evolution of materials under different external stimuli with high spatial and temporal resolution. The common theme of these techniques is that nowadays the experimental data is often represented as a three- or higher-dimensional data set as shown in Fig. 1 (left) [2].

The ever-growing data complexity, size, and speed at which it is created in experiment renders human-based analysis not only impractical, but also largely limits the discovery of latent features, which often equip a material with a certain functionality [2]. This has stimulated the development of automated computer-based and machine-learning analysis algorithms to harvest the rich information contained in the data and to turn the data into interpretable physical quantities [1]. For example, principle component analysis and clustering were employed to automatically separate different phases in a bismuth ferrite sample at atomic resolution obtained from a multi-gigabyte 4D-STEM data set [3]. The development of open-source-based data-analysis tools has been paramount for treating and interpreting multidimensional and large-scale data sets from different microscope manufacturers in an efficient manner and provide flexible platforms towards on-the-fly data analysis even of big data sets [4].

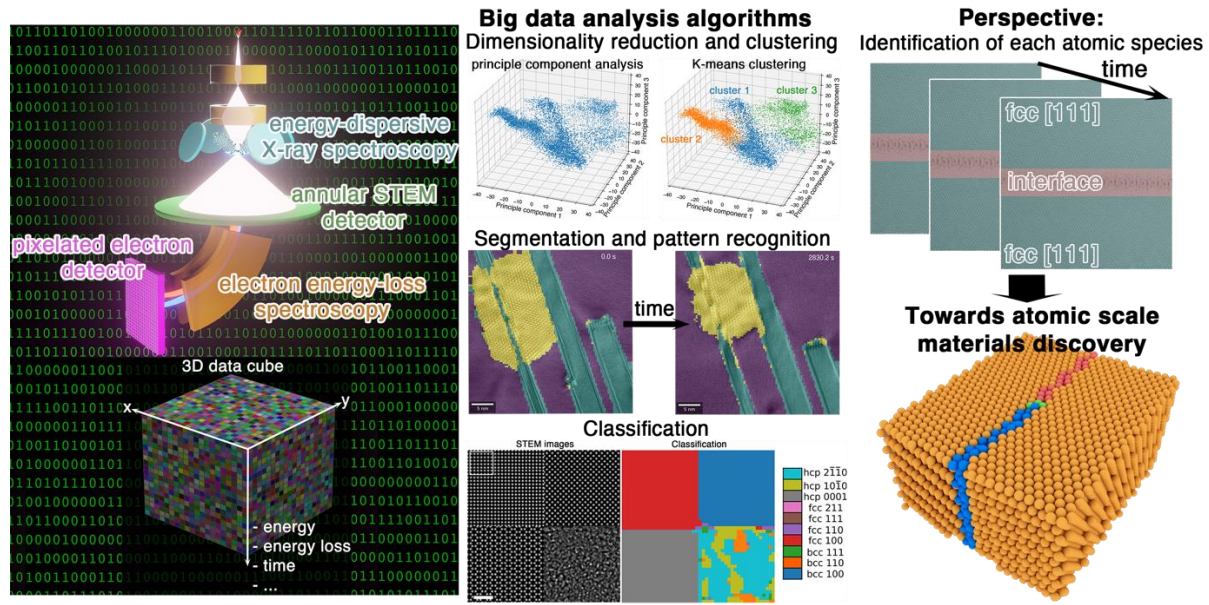

**Figure 1** Multimodal data streams and related high-dimensional data representation (left). Data-analysis algorithms for dimension reduction of large-scale data, automatic pattern recognition / segmentation and quantitative classification of the data (middle). Perspective to harvest the variety of signals and content contained in big data to uniquely identify the 3D physical structure of a sample with atomic precision, its evolution with high time resolution to discover new material phenomena on the atomic and electronic scale. Partially reproduced from [5], https://creativecommons.org/licenses/by/4.0/.

**Current and Future Challenges**

Incremental data acquisition and analysis are still common even in modern microscopy laboratories. The experiment is sequentially followed by the interpretation of the collected signals and eventually the loop repeats with refined measurements until sufficient insights into physical material quantities are gained. There are several challenges associated with this incremental approach in the era of big microscopy data:

1.  handling, storage, and labeling of the data to enable reproducible data analysis
2.  human-based data analysis often largely exceeds experimental time frames
3.  limited interdigitation of data acquisition and analysis
4.  lack of automated or autonomous data analysis tools

These technical restrictions often directly compromise material characterization and with this new material discoveries. One of the greatest challenges is the interdigitation of data-stream generation in a microscopy experiment and its direct analysis to provide live feedback to the researcher. Different approaches can be envisioned here where parallelized high-performance computation (HPC) utilizing modern graphical processing unit (GPU) capabilities is directly performed at the microscope computer [6] or edge computing in a distributed system, where the HPC tasks are performed either on cloud servers or at HPC centres [7].

The broad variety of data streams utilized to probe materials ranging from simple 2D images, to 3D or higher dimensional hyperspectral data sets, to time series probing material evolution or 3D tomographic reconstructions require the development of versatile and autonomous data analysis algorithms. Typically, advanced algorithms to reduce the dimensionality of hyperspectral data, segment or recognize patterns in images, and classify features in multidimensional data sets are employed as separate or sequential instances as shown in Fig. 1 (middle) [1], [2]. It has been shown that unsupervised machine learning is capable to automatically segment different crystalline regions in atomic resolution images and video sequences solely based on crystal structure symmetry without requiring prior knowledge on the underlying structure [8]. Using a trained Bayesian deep neural network , it is even possible to classify crystal structures in atomically resolved images and identify defective regions or interfaces by considering the uncertainty in the prediction [5]. In a future direction, one would envision that novel big-data and machine-learning algorithms will be integrated in hybrid algorithm architectures that perform automatic or even autonomous tasks.

**Advances in Science and Technology to Meet Challenges**

Advances in computing architectures for microscope laboratories are one side of the coin, but integrated or hybrid machine learning based algorithms need to be deployed alongside to enable automatic analysis of large-scale data. Recent developments in machine-learning and in particular deep-learning approaches in electron microscopy hold great promise for laying the foundation for autonomous data-analysis and electron-microscope operation [1], [9]. Ultimately, the aim is to enable the discovery of new material phenomena and to probe the physical properties of materials and their evolution with atomic precision. Since the physical nature of electron wave propagation and interaction in a crystalline material is well understood, ground truth training data for a deep learning model can be efficiently generated [10]. However, a large deep-learning model would need to contain information not only of all known crystal structures and phases, but more importantly of different point, line, or planar defect types. Recognizing defects from supervised learning, however, is nearly impossible to achieve at the day of writing, since the atomic configurations existing in nature are not necessarily known or understood. Instead, a convolutional neural network can be trained on simulated images of pristine crystal structures, while still localizing and obtaining information on material imperfections [5].

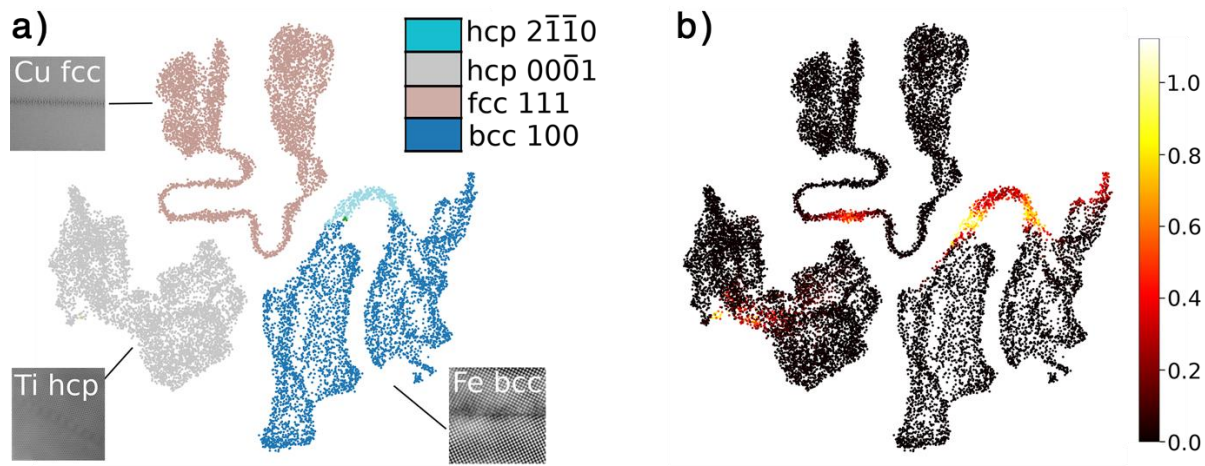

**Figure 2** Dimension reduction of neural-network representations of a classification model trained on simulatd atomic resolution STEM images of pristine crystal structures. Each point in the scatter plots corresponds to a local image patch of an experimental image. The color scale corresponds to two items of information that the model provides: a) Classification assignments of experimental STEM images of interfaces, here grain boundaries, in face-centered cubic (fcc) Cu, body-centered cubic (bcc) Fe and hexagonal close packed (hcp) Ti. b) Mutual information quantifying the uncertainty of the prediction of the deep learning model. Bulk regions appear as clusters of low

model uncertainty while interfaces correspond to diluted regions with high model uncertainty. Reproduced from [5], https://creativecommons.org/licenses/by/4.0/.

Figure 2 shows the neural network representations obtained after dimension reduction of the fully connected layer before the classification and the corresponding uncertainty of the prediction. Although the model was trained on pristine crystal structures, it is capable to distinguish the different types of interfaces (here: grain boundaries) and the model uncertainty provides an indirect way to locate material imperfections. Approaches combining supervised, unsupervised and active learning are needed to further explore regions in data sets with high uncertainty, which may represent an unknown interface structure or surface configuration. Furthermore, the classification tasks have to be extended to also consider local composition and electronic structure to fully exploit the data and yield a holistic picture of the physical nature of a material on the atomic level. Future models should enable live feedback at high time resolution to facilitate autonomous steering of the experiment and consider active re-training to include disturbed or unknown atomic structures.

**Concluding Remarks**

Big data in electron microscopy is already a reality and will play an increasing role in the future not only for the sake of data acquisition, but to holistically characterize every single atom in a material paving the way for atomic scale materials discovery. Spectroscopic and scanning diffraction data sets (e.g. 4D-STEM) contain information on the elemental nature, the electronic and 3D structure of a material and hence this information needs to be fully harvested. Technological advancements in computing infrastructure have to be developed in parallel with hybrid machine learning algorithms in electron microscopy laboratories to move away from incremental experimentation. Combinations of unsupervised and supervised learning approaches have the potential to automatically identify and label different crystal structures and atomic species in complex data sets and will eventually uncover latent patterns in an automatic fashion. This will guide scientists to interesting regions in a sample and accelerates the deployment of physical material models.

**Acknowledgements**

The authors acknowledge founding by BiGmax, the Max Planck Society's Research Network on Big-Data-Driven Materials Science.

### 4.3: Applications of Machine Learning in the Acquisition of and Knowledge Extraction from Experimental Data with a Focus on Electron Microscopy

Marcel Schloz[1], Anton Gladyshev[1], Meng Zhao[1], Thomas Kosch[2], Christoph T. Koch[1]

[1] Humboldt-Universität zu Berlin, Department of Physics, Berlin, Germany
[2] Humboldt-Universität zu Berlin, Department of Computer Science, Berlin, Germany

**Status**

The success of inherently big-data-based machine learning (ML) in materials science can also be observed in its sub-field of structure research through (scanning) transmission electron microscopy, (S)TEM [1,2]. Here, ML has become a game changer for post-acquisition data analysis, such as image reconstruction [3], improvement of data by denoising and resolution enhancement [1,2,4] and structure recognition [1,2,5,6,7]. One of the bottle-necks for the efficiency with which electron microscopes can generate materials knowledge is also the investment of time and human, highly microscope-specific expertise required to align the instrument for optimal performance, especially, when the materials question to be solved requires switching between different modes of operation. While some big-data-trained ML models have already demonstrated to be capable of measuring aberrations very quickly [8], they are not yet capable of handling the complexity of a modern microscope which, for some instruments, requires managing more than 500 current supplies. A few groups are also applying ML methods towards real-time data analysis and automating experiments as illustrated in Fig. 1 for the case of STEM [9]. In contrast to the field of cryo-electron microscopy, where fully automated experiments can run for multiple days by repeating the same image acquisition process for automatically exchanged samples at many pre-defined sample positions, the complexity of adaptive ML-driven experiments in materials science (S)TEM experiments is much higher, given the inhomogeneity of most samples, the wide variety of signals to choose from and switch between, and the sequential process with which the data is acquired. Conventional ML methods used on already acquired data sets can simply not be applied one-to-one. New ML approaches for real-time applications in electron microscopy are still rare and most notably their on-the-fly implementation on the microscope has so far not been realized [9].

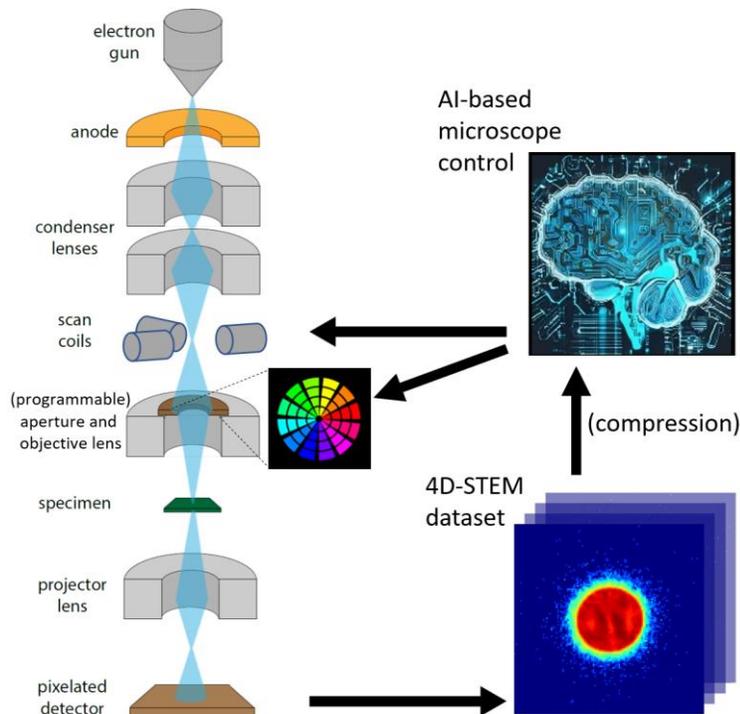

**Figure 1** Schematic of an AI-controlled scanning transmission electron microscope using a pixelated detector to acquire 4D-STEM datasets. Electrons are emitted from the electron gun and guided through a system of electromagnetic lenses and are deflected by scan coils before they interact with the specimen. The electron beam is then guided to the detector, which records a diffraction pattern of approx. 1 MB size for thousands of scan position, thus, resulting in a 4D-STEM dataset of typically 10s of GB in uncompressed size. ML methods analyse the raw or compressed data in real-time and control the hardware components of the microscope to optimize the experiment. The controlled components shown here are the scan coils and a programmable phase plate (inspired by the commercially available design by adaptem.eu), but it can also be lens currents, aberration-corrector settings, etc.

### Current and Future Challenges

In addition to the requirement for very fast data processing and fast access to electron optical components of the microscope, method developments will also need to consider the following two crucial components: The first key component is the fast handling of huge microscopy data. Electron microscopes can nowadays acquire several GBs of data within seconds which, means that ML methods for real-time applications should be capable of processing huge data sets within a fraction of a second. Obviously, a tight integration between hardware and software will play a crucial part in the solution to this problem. Edge computing and camera integrated compression techniques [10] are here just two examples to be mentioned. Another important component for the development of new real-time ML methods is a high level of adaptability. The environment in the microscope constantly changes between, but sometimes even during experimental sessions. ML methods need to deal in real-time with data that has been acquired under these circumstances without a significant loss in performance. Furthermore, methods that aim for an automation of the experiment are required to easily adapt to different experimental goals.

The high complexity and cost of ownership of state-of-the-art electron microscopes allows only a few labs staffed with expert operators who have undergone extensive microcope-specific training to run them. Maximizing these instrument's scientific output per time as well as democratizing access to them calls for improving their user interface in analogy to how modern chatbots have recently started to enable anybody to write complex computer programs.

**Advances in Science and Technology to Meet Challenges**
Advances in the method development that combines deep learning and reinforcement learning (RL) show promise that dynamic decision-making problems can be solved with a strong performance by a machine alone. Operating an electron microscope in an automated fashion could therefore benefit from this development. A first step towards this direction has been proposed in Ref. [9], where the combination of deep learning and RL offers the possibility to perform low-dose experiments for electron ptychography through adaptive scanning. A schematic of the adaptive scanning workflow is shown in Figure 2. The advantage of this approach is that it is highly adaptable to a wide range of scanning microscopy techniques through the modification of a reward function that expresses the research goal. Hence, various imaging and spectroscopy techniques, such as STEM EELS, that have already been shown to benefit from an optimized scanning scheme, could be further advanced through a successful automation of the experiment. But also many other parameters of the experiment, such as adjustable aberrations, lens currents, or the phase shifts in programmable phase plates (see Fig. 1) can be optimized to improve the efficiency of the experiment with which a given research question is being addressed. Recent developments in software for processing natural language are likely to result in the highly technical user interface of electron microscopes being extended by chatbots and comparable features.

In order to deal with 10s of GB of data per scan, it has been shown [10] that compression based on data-dependent linear transformations yields superior results when compared to conventional techniques like binning, or singular value decomposition, both in terms of compression ratio and quality of the information extractable from the data-set. The integration of artificial neural network (ANN)-based feature recognition techniques has the potential to further enhance compression performance. Before collecting the main data-set, a network can be pre-trained in a similar way to adaptive scanning [9], but with the aim of capturing the diffraction patterns as best as possible with as few values as possible.

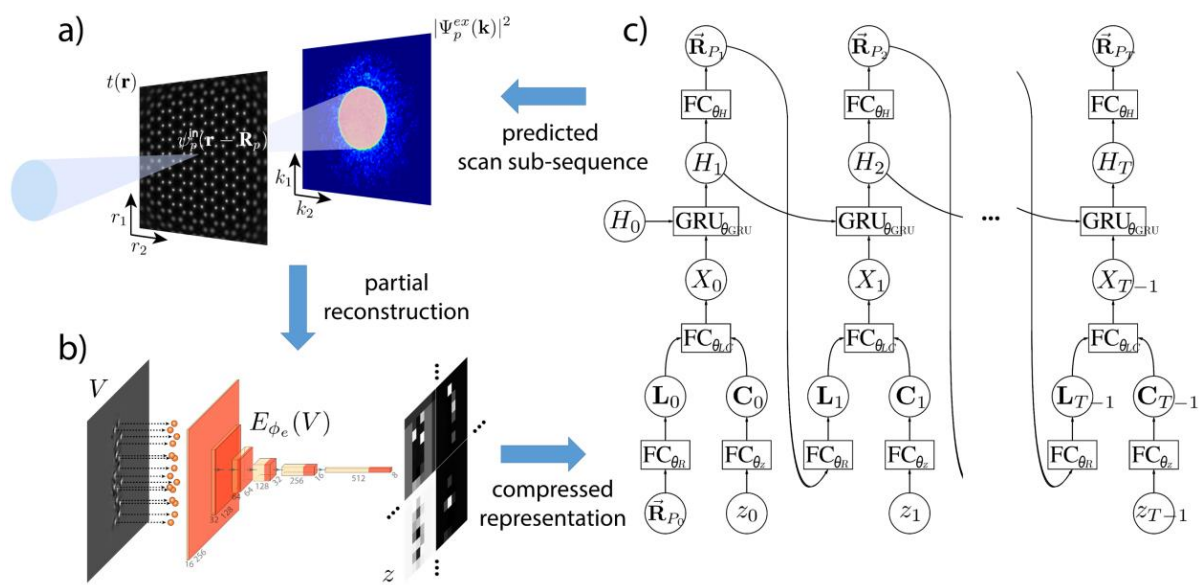

**Figure 2** Schematic of a ML-driven adaptive scanning workflow for the purpose of optimizing the scanning in a 4D-STEM experiment in real-time. The employed ML methods consist of a convolutional neural network for the atomic structure extraction and a recurrent neural network for the sequential prediction of scan positions. Training of the networks is performed through RL. Reproduced with permission from Springer Nature [9].

## Concluding Remarks

In summary, big-data-based ML methods have already shown to be very powerful for post-processing tasks of electron microscopy data, but given the high complexity of these microscopes, their application in useful real-time data analysis and experiment automation methods still lags behind. Some initial developments of workflows that leverage ML methods to perform and optimize specific tasks of an electron microscope show promise for transitioning this fully human-controlled instrument to a (partially) autonomously operating machine being capable of carrying out precision measurements in a fully documented and fully reproducible manner. We expect that this development will largely increase the research output obtained from this type of instrumentation.

## Acknowledgements

M.S. and C.T.K. acknowledge financial support from the DFG - Project-ID 414984028 - SFB 1404.

## 4.4: Machine Learning for Analyzing Atom Probe Tomography Data

Yue Li [1], Ye Wei [2], Alaukik Saxena [1], Christoph Freysoldt [1], Baptiste Gault [1,3]

[1] Max-Planck-Institut für Eisenforschung GmbH, Düsseldorf, Germany
[2] Ecole Polytechnique Fédérale de Lausanne, School of engineering, Lausanne, Switzerland
[3] Department of Materials, Imperial College, London, UK

**Status**

Atom probe tomography (APT) is a burgeoning characterization technique that provides compositional mapping of materials in three-dimensions at the near-atomic scale [1]. The data obtained by APT takes the form of a mass spectrum, from which the composition of the analysed material can be extracted, and a point cloud that reflects the distribution of all the elements within the region-of-interest of the material being studied. Material-relevant data must be extracted from this point cloud through the use of data processing or mining techniques. These go from simply the local composition of a phase or a microstructural object, sometimes extracted via cluster-finding or nearest-neighbour algorithms today classified as machine-learning but used in the APT community for many decades [3]. Phase morphology or even partial structural information can be obtained but the information can be limited or distorted because of trajectory aberrations that are caused by heterogenities in the specimen's end shape down to the near-atomic scale. Today, data reconstruction and processing is most often done in commercially-available software, which does not allow for exploiting the cutting-edge methods arising from big data and machine-learning, and also remains very much user-depedent [4]. The enormous potential to mine atom probe data is clear, but this requires complete FAIR-compliant analysis workflows that make use of machine-learning to facilitate more reliable and reproducible data processing and extraction, to really go beyond what human users can achieve. This section reviews challenges of APT data analysis (partially) solved by the application of machine learning and points out the remaining crucial locks to be addressed in the future.

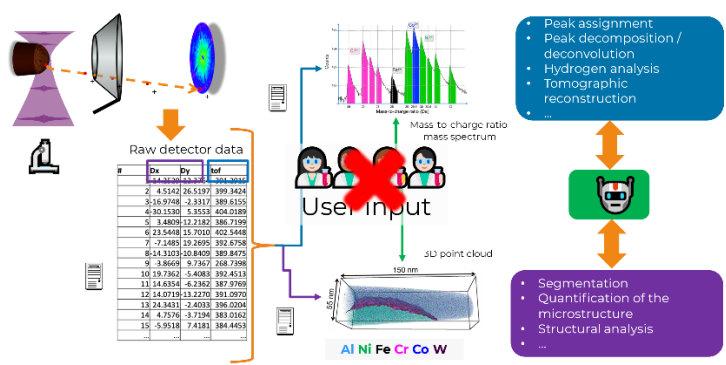

**Figure 1** summary of typical APT data analysis workflow, from the processing of the experimental data to form and analyse the mass spectrum to the reconstruction of the 3D point cloud, all of these steps typically require user input, highlighting potential for ML-learning and possible complete data processing workflows.

**Current and Future Challenges**

A critical challenge is that present-day APT data processing tools and workflows are inherited from "traditional" interactive data analysis based on user-interactions, through a fixed set of data analysis techniques and visualization that leaves little flexibility to explore novel and processes, as summarised in Figure 1. User-input includes assignment of peaks to particular atomic or molecular species to manually retrieved structural information and microstructure segmentation and quantification. Machine-learning has the potential to automate many of these analysis steps, with models that are based on physical input and constraints. Some progress has been made across the community with dedicated machine learning algorithms to mine compositional [5] and structural information [6]. For instance, for mass peak assignment, we introduced an approach that uses known isotopic abundances to identify patterns in mass spectra, outperforming human users without loss of accuracy [7]. Following reconstruction of the 3D point clouds, automated identification and quantification of grain boundaries were proposed, and for more general microstructure segmentation, Saxena et al. [8] introduced an approach that uses clustering in the compositional space, demonstrating unique capabilities for segmentation of the various phases, along with the quantification of their composition and morphologies. These would normally have been extracted through manually positioned regions of interest, which is time-consuming and error-prone. Structural imaging by APT is hindered by the anisotropic spatial resolution and the limited detection efficiency [3]. Recent efforts have managed to overcome these for the ever-challenging analysis of chemical short-range order (CSRO) by using convolutional neural networks, using the workflow in Figure 2 [9]. A key challenge for the future is to move away from the developing individual tools to tackle isolated problems to think about complete data analysis workflows, from patches to a logical patchwork that will also facilitate adoption. A way to solve this would be to open the programs themselves via APIs at all levels, or at least facilitate data exchange through open data formats accessible to external processing by independent tools.

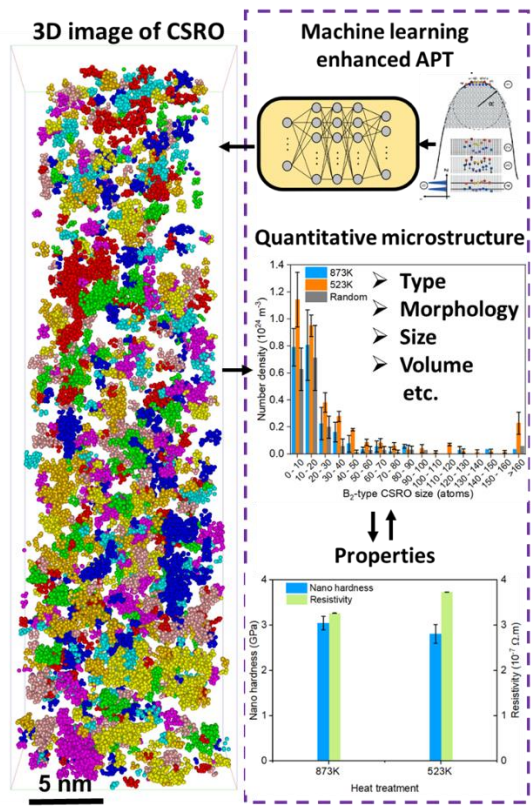

**Figure 2**. Machine-learning enhanced APT to break the inherent resolution limitation of atom probe tomography, and precisely image multiple arrangements of atoms associated with CSRO, in 3D [5].

**Advances in Science and Technology to Meet Challenges**

For APT, post-processing is mostly executed with proprietary software tools, which can often be opaque in their execution and have often limited performance and preclude the facile deployment of novel data processing methods. There is a need to agree on a more opened data format and metadata conventions as a critical prerequisite. Development of machine-learning optimized hardware and software remains plagued by the use of proprietary, specific data format, which limits usage across software, techniques and communities. And this should include the raw data, not only what has already been processed. As such, the community will have push to provide a complete set of tools equivalent to the currently available integrated beginning-to-end workflows, i.e., from an experiment to a publishable image, yet these will have to be open and extendable to include machine-learning steps and fully documented to also include traceable information regarding the sample and the specimen with appropriate metadata. A prerequisite is also the use of open and documented data formats. As a preliminary efforts in this directions, let us mention here Paraprobe [10], that is fully open-source and provides clear documentation of each analysis step for post-processing APT datasets that offers orders of magnitude performance gain, automation, and reproducibility. For now, these open tools are seldom used, and the community seems to wait user-friendly platforms, which so far do not exist. This hinders complete FAIR-workflows that are so far lacking, which precludes direct correlations with other computational or experimental techniques, but also wider meta-analyses as introduced by Meier et al. [5]. Finally, there is a need for a repository of benchmark datasets that would allow to evaluate the performance of new developments in a transparent way across the community.

**Concluding Remarks**

Although the above-mentioned efforts have demonstrated the potential for state-of-the-art machine learning to meet existing challenges for APT data processing, some major aspects remain to be tackled to fulfil the full potential. Machine-learning has the potential to help address many of APT shortcomings, and, for instance, resolve some aberrations that plague the accuracy of the measurements by better interfacing with modelling efforts in APT. This is necessary to reach true atomic-resolution that will help extract more precise local atomic arrangements. Optimisation of the data acquisition, and establishing a dialog between the instrument and the data processing are also areas that will need exploring – in this regard, APT is far behind other high end microscopy techniques. Overall, we are only at the beginning of the use of machine-learning for APT, but the preliminary work that has been done across the community lays solid ground to build better, more encompassing and efficient tools in the future.

**Acknowledgements**

Y. Li acknowledges the research fellowship provided by the Alexander von Humboldt Foundation. A.S. appreciates funding by Helmholtz School for Data Science in Life, Earth and Energy (HDS-LEE). BG acknowledges support from the Deutsche Forschungsgemeinschaft (DFG) for funding from the Leibniz Prize 2020 (GA 2450/2-1) and for funding of the TRR 270 HoMMage (INST 163/578-1). BG, YW, YL and CF are grateful for financial support from BiGmax, the Max Planck Society's Research Network on Big-Data-Driven Materials Science. The colleagues from the Max Planck Computing and Data Facility, Garching, Germany are warmly acknowledged for their help and support.

## 4.5: Data-Driven Approaches for Heterogeneous Catalysis

Andrew J. Logsdail, [1,2], C. Richard A. Catlow, [1,2,3,4], Lara Kabalan, [1,2,5], Igor Kowalec, [1,2] and Zhongwei Lu, [1,2]

[1] Max Planck Centre on the Fundamentals of Heterogeneous Catalysis (FUNCAT), School of Chemistry, Cardiff University, UK
[2] Cardiff Catalysis Institute, School of Chemistry, Cardiff University, Cardiff, UK
[3] UK Catalysis Hub, Research Complex at Harwell, RAL, Oxford, UK
[4] Department of Chemistry, University College London, London, UK
[5] STFC Hartree Centre, Daresbury Laboratory, Daresbury, Warrington, UK

**Status**

Heterogeneous catalysis is vital to sustain humanity and to address important societal challenges such as achieving net zero. Heterogeneous catalysis is also chemically complex, and the realisation of new catalysts is challenging. Catalysts themselves can contain multiple active elements; for example, the catalyst used for the Haber-Bosch process, which is integral to feeding 50% of the global population, typically contains iron, aluminium, calcium, potassium, and oxygen, with activity subtly dependent on composition. [1,2] The composition of catalytic materials can be explored successfully via data-driven approaches, yet catalytic reactions occur at the surfaces and interfaces of these materials, and therefore the material properties must be investigated also as a function of the reactive surfaces and interacting medium; [3] furthermore, a rational design process must also consider the intricacy of reaction mechanisms to ensure appropriate reactivity and product selectivity, which includes sensitivity to temperature and pressure, to result in truly industrially relevant catalysts. The complexity of such catalytic systems quickly becomes intractable to fully explore with current experimental or computational efforts.

Historically, catalysts have been identified and their application optimised via empirical investigations, using previous success to guide future decision-making. Such "top-down" experimentation has recently seen the integration of high-throughput experimentation (HTE) into workflows, accelerating catalyst discovery through parallelisation of testing; in the more advanced cases, the HTE is coupled with data-driven analysis of reactivity/selectivity and automation to self-consistently optimise the efficacy of the catalytic system towards a target property, working within a defined parameter space. [4] The current HTE approaches do not typically include advanced *in situ* or *operando* characterisation, but these methods are increasingly available separately and benefit from similar emergent capabilities in automated data-driven analysis.

Alongside experiment, the advancement of computational capabilities allows the "bottom-up" interrogation of elemental and structural knowledge from across the periodic table, presenting significant opportunities for accelerated data-driven discovery. Promising materials can be considered further using parameterised models to explore surface structures and composition as a function of operating conditions, [5,6] and reaction mechanisms derived using automated construction of chemical reaction networks, providing vast quantities of data relating to a reaction landscape [7] from which rates and product distributions are accessible via kinetic modelling. With the knowledge calculated within this sampling space, the efficacy of the catalytic system can be linked against key "descriptors" of the catalyst and its operating conditions, providing powerful shortcuts when navigating across the reaction landscape to find better catalysts via e.g., active learning protocols. In the most state-of-the-art approaches, descriptors are derived as compound functions of both experimental and computational information, via multi-fidelity data models. [8]

**Current and Future Challenges**

Data-driven models are dependent on large, accurate, and complete datasets, yet such experimental data remains challenging to locate, access, use, or reproduce. Historically, the reward structure of the research community has been towards positive results, which means that negative results are not shared. [9] Incomplete datasets lead to sampling bias and inaccuracy of data models; furthermore, hidden data can also present a challenge for reproducibility, whereby not all the experiment parameters are reported for future investigators. Data quantity and quality are also important aspects, yet most experimental investigations typical focus in a small chemical space, which lead to small datasets. Indeed, data completeness can again become challenging when only the "best" catalysts are considered for higher-level characterisation methods, such as *in situ* electron microscopy; and simultaneously, data quality is compromised, as differing standards of analysis are introduced, and outcomes reported in contrasting formats. [10] The combination of identifiable data sources is also a current challenge, as the quality and quantity of information can vary in relation to synthetic methods, catalytic testing, and characterisation; and these data may be embedded in images, making collecting accurate data a challenge.

Similar challenges relating to data completeness and accuracy exist in the computational catalysis domain. Here, greater efforts have been made to creating standardised, complete, and publicly available datasets, [11,12] yet the realisations often remain limited to subsets of catalysts/reactants/products (e.g. oxygen evolution reaction electrocatalysts [12]) and a current challenge is to expand knowledge space. More pertinent is the need for accurate computational data that can be confidently correlated with experiment. Considering machine-learning forcefields (MLFFs), which are a notable success from the application of data-driven approaches in materials modelling, a current challenge is to build these approaches to reproduce *experiment*, and not just higher-level computational models. Further extension of the MLFFs should then include multiple compositional and environmental aspects of a fully operational heterogeneous catalyst; and for this more efficient modelling paradigms are needed to create bigger datasets. Future challenges then arise with the integration of computational and experimental datasets, whereby parameters and observables from each respective domain must be collated and compared on an equal footing to provide value to the researchers of the future.

**Advances in Science and Technology to Meet Challenges**

There are multiple technological advances identifiable to meet the challenges and fully achieve the potential of data-driven approaches. Within the laboratory, greater accessibility of automated high-throughput facilities, capable of synthesising, testing, and characterising catalysts, will be powerful in facilitating on-the-fly data-driven catalyst discovery, and must be coupled with public accessibility in centralised repositories to achieve larger, consistent, and more complete datasets. For modelling, improved software models are still needed to simulate a more accurate description of complete catalytic conditions, including the effects of temperature, pressure, and solvents, to provide accurate surrogate models of energy landscapes that can be explored rapidly, with automated discovery again an opportunity. And at the interface of computation and experiment, greater integration of catalytic datasets to provide holistic coverage is necessary to account for deficits in knowledge from either the experimental or computational domains alone; indeed, one needs to harness the individual strengths of "top-down" and "bottom-up" perspectives to derive complementary data, rather than distinct.

These scientific and technological advances are coupled also with a need for greater discussion between members of the catalytic community, and advocacy of standardisation. Whilst the principles of findable, accessible, interoperable, and reusable (FAIR) data have developed strong roots in the computational modelling domain, the distribution or centralisation of experimental data remains limited, and focused on positive results. The value of *all* data should be championed, and the importance of metadata to aid users in understanding value and limitations of a given dataset; deposition of results in an accessible resource should be encouraged, especially for experiment, where uptake is more urgently needed. The communication between researchers should include experimental and computational communities, and span academia, industry, and third-party organisations, at all levels of scientific investigation, in order to deliver better understanding of data needs and standardisation of data-collection procedures. The work here is implicitly multidiscipline, and so the interaction of chemists, materials scientists, physicists, computer scientists, data scientists and other domain experts should be encouraged to maximise the opportunity for multi-fidelity models that address shortcomings arising in individual research domains. Finally, there is the need to train and distribute knowledge among researchers of the value of their data; we should be educating in a cross-disciplinary manner about the importance of detailed digital data collection, in both experiment and computation. Such action will lead to engagement and investment towards necessary tools to accelerate the big-data driven discovery in heterogeneous catalysis; such software capabilities already exist, driven by the explosion in interest towards data-driven discovery, but the potential is yet to be realised.

**Concluding Remarks**

The status for data-driven approaches in heterogeneous catalysis is promising, with strong application in computational fields and increasing demonstrations of potential in experimental laboratories. However, challenges remain with respect to ensuring the quality and completeness of individual datasets, as well as improving accessibility and standardisation. Opportunities have been highlighted that include increased automation within research environments, improved cross-discipline communication, and efforts among users to reach distribution standards that will benefit emergent as well as established researchers. Catalysis is an extremely challenging but valuable field, with impact on all of humanity. Adoption of the outlined approaches can facilitate the update of emergent data-driven methods for a transition to cleaner, more active heterogeneous catalysts that benefit the global population. There are many examples of good practice, but efforts are still needed if we are to maximise the potential value for all.

**Acknowledgements**

The authors are grateful for funding by the EPSRC Centre-to Centre Project (Grant reference: EP/S030468/1). AJL acknowledges funding by the UKRI Future Leaders Fellowship program (MR/T018372/1). ZL acknowledges funding by the China Scholarship Council.

## 4.6: Synchrotron Small Angle X-Ray Scattering – Perspectives of Machine Learning

Peter Fratzl[1]

[1] Max Planck Institute of Colloids and Interfaces, Potsdam, Germany

**Status**

X-ray scattering and diffraction pertain to a major set of techniques to characterize the structure of materials at the nanoscale. Small-angle x-ray scattering (SAXS), in particular, has been developed in the 1950s to resolve structures in the size range 1 – 100 nanometers [1]. Despite the development of electron microscopes some years later, it remained an important technique, mostly because x-rays are less strongly absorbed than electrons, which allows for in-operando experiments, studying the effect of physical stimuli, such as temperature, pH or humidity on material structure. A strong boost in the use of small-angle scattering came with the availability of synchrotron radiation that improved the time resolution of in-operando experiments, but also opened to possibility to transform SAXS into a multiscale imaging tool. In this approach, the general idea is that nanoscale information is extracted from analyzing the scattering patterns, while mapping of the specimens provides the information at the microscale (see Fig. 1). The first attempts with SAXS-based imaging go back to the 1990s [2]. This evolved until the development of SAXS tomography which yields six-dimensional data: three dimensions in real space through scanning and rotating the specimen (typically with micrometer resolution) , as well as three additional dimensions from the scattering patterns within each voxel (containing nanoscale information) [3, 4].

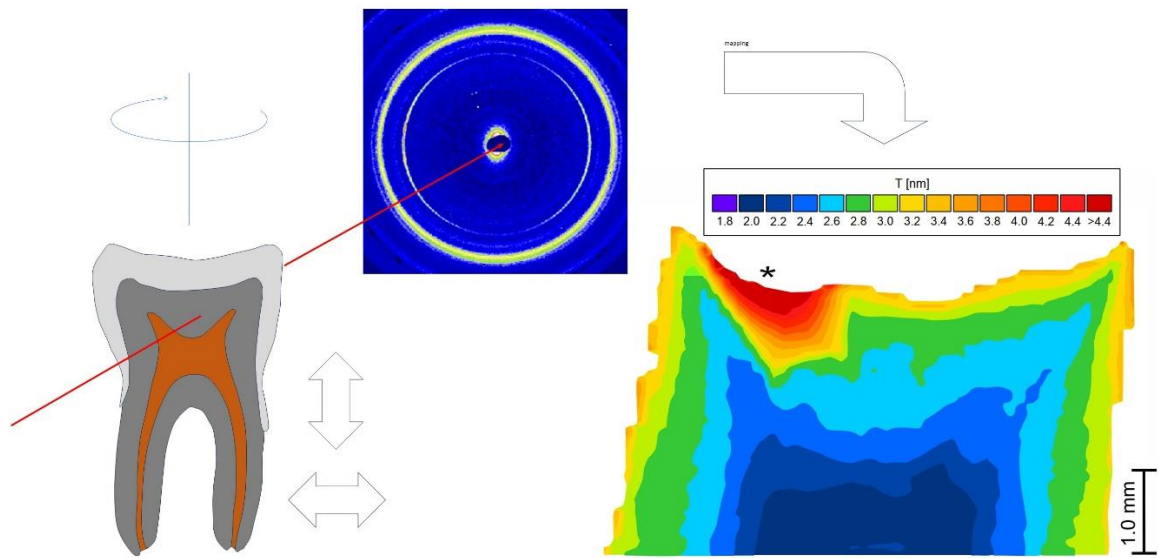

Fig. 1: Principle of scanning-SAXS imaging. The specimen (for example a tooth section) is scanned across the x-ray beam with a diameter between tens of nanometers and several micrometers. Parameters extracted from the scattering patterns can then be mapped with a resolution corresponding to the x-ray beam diameter. In the figure, this is the thickness of mineral particles in dentin (the star indicates an area with a caries lesion). Picture adapted from [5].

The enormous advance in the brilliance of x-ray beams, as well as in x-ray optics enables not only the collection of multidimensional SAXS-tomography data but also the measurement of massive numbers of specimens even within short times.

**Current and Future Challenges**

These advances upstream of the specimen in the experiment, however, lead to new challenges downstream of the specimen, linked to the treatment and the evaluation of massive amounts of data. A schematic of the workflow in a SAXS measurement is shown in Fig. 2. The traditional way of conducting such an experiment would be the path symbolized by (A) and (B) in this figure. (A) represents specimen preparation and the experiment planning and (B) the data collection. These data would then be brought back from the synchrotron experiment for treatment and analysis. However, with the increased speed of data collection, a general challenge in this approach resides in the fact that the experimentalist is essentially blind without some capabilities of data diagnostics. This requires elementary pre-analysis of the data to see whether a modification of the beamline setup could improve the experiment. Recognizing this, software packages involving fast data diagnostics were developed, an example being DPDAK, an open code software introduced at the BESSY and the DESY synchrotrons (in Berlin and Hamburg, respectively) [6].

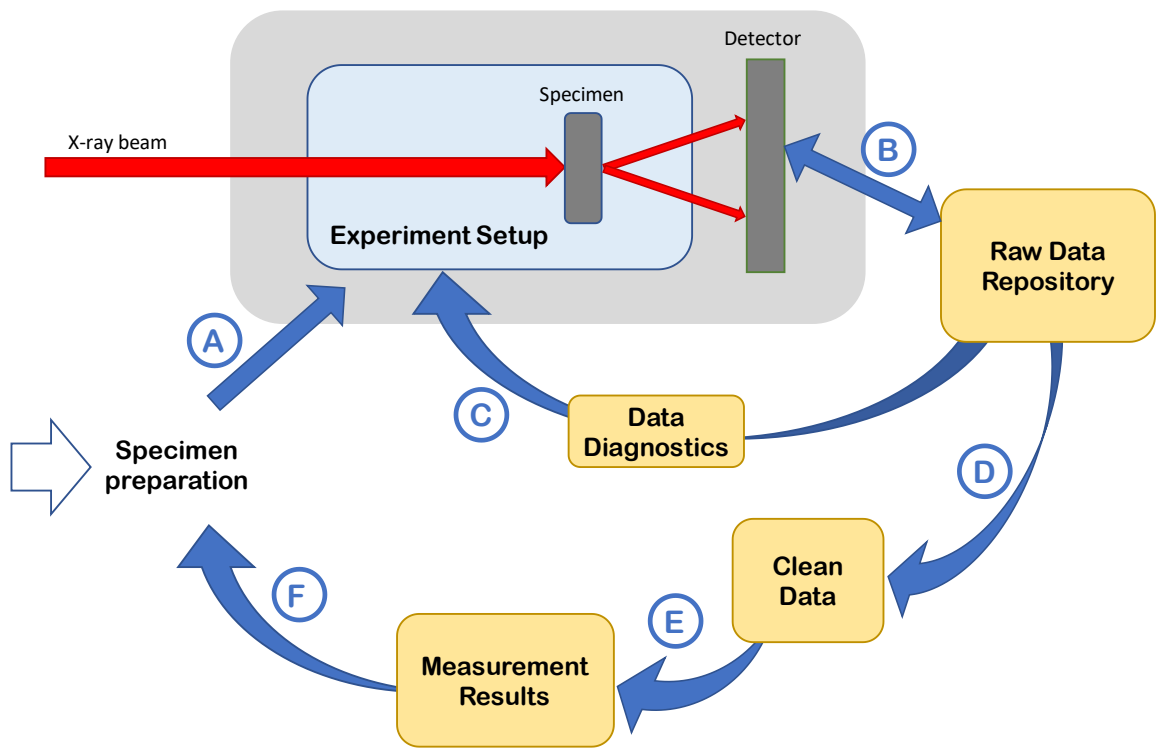

Fig. 2: Schematic workflow of a small-angle x-ray scattering experiment. The traditional approach would be characterized by the arrows A (experiment planning to define the experiment setup) and B (data collection). With increasing data rates, a number of feedback loops involving machine learning are beginning to improve the quality and speed of the experiment: C is a readjustment of the experiment setup based on rapid data diagnostics, D is data reduction and denoising, E is data analysis and F automatic material synthesis based on the measurement results.

With the amount of data collected in each beamtime session increasing continuously over the years, a number of additional challenges appear from the fact that manual data treatment becomes impossible. This applies to the cleaning of data (such as denoising, background subtraction, image reconstruction, normalization, etc.) and even more to the data analysis, which in SAXS often involves data fitting. These steps are indicated by the arrows (D) and (E) in Fig. 2.

**Advances in Science and Technology to Meet Challenges**

Especially in SAXS tomography experiments, radiation damage should not be underestimated., since every specimen position will be hit several times by an intensive x-ray beam due to the required rotation of the specimen around multiple axes [4]. A typical strategy is then to reduce the irradiation time, which inevitably increases the noise in the data. To avoid problems with this noise in the 6D data reconstruction after the measurements, Zhou and coworkers propose a machine learning (ML) for the denoising of scattering data [7]. This approach facilitates step (D) in the diagram of Fig. 2.

The reconstruction of SAXS tomography data is equally challenging due to their high dimensionality. A possible traditional approach consists in calculating invariants of the SAXS data before reconstruction, which replaces the three-dimensional SAXS data by scalars that can be reconstructed much more efficiently [8]. SAXS invariants are useful, since they contain information about volume and surface of nano-size objects in the specimen [1] and allow, for example, the calculation of particle sizes in bone or dentin [2, 5, 8]. In the last few years, ML approaches are being developed for tomographic data reconstruction. Omori and coworkers review these developments for tomography using SAXS but also x-ray diffraction and other modalities [9]. While these advances relate to step (D) in Fig. 2, the review also addresses ML approaches for segmentation and analysis of the reconstructed data [9] (step (E) in Fig. 2).

Once data are reconstructed, every voxel in SAXS tomography data contains a scattering pattern to be analyzed. This means a massive effort for data analysis (Step (E) in Fig. 2) after reconstruction. Similar numbers of SAXS patterns need to be analyzed in other situations, for example when material structures are studied as function of physical parameters (temperature, pressure, pH, humidity, etc.) in multiple measurements. A recent review by Anker and coworkers addresses machine learning (ML) approaches to analyze a range of synchrotron-based experiment data, including SAXS but also powder diffraction, pair distribution function, inelastic neutron scattering and X-ray absorption spectroscopy data. While the traditional approach would be to fit a physical model to the data, supervised ML can be used to train a model for the prediction of structure based on data, but also to predict the scattering data based on a known structure and also to predict parameters based on some physical understanding of the system [10]. In another recent work [11], a ML-based analysis of SAXS data is proposed, which is based on Gaussian random fields that avoids the common model fitting of the data.

The approaches discussed until now are improving workflows in nearly all steps of SAXS experimentation (step (C) to (E) in Fig. 2). A last step (F) potentially closes the loop towards a fully automatized experimentation. This challenge is currently been taken up under the label of Autonomous Experimentation. Beaucage and Martin report on the development of an open liquid handling platform for autonomous formulation and x-ray scattering [12]. Yager and coworkers review this new paradigm and show how autonomous x-ray scattering can enhance efficiency, and help discover new materials [13]. **Concluding Remarks**

Small-angle x-ray scattering is an old method that is currently seeing an enormous increase in activity due to highly brilliant x-ray sources, more performant x-ray optics and – most recently – rapid progress in the treatment and the analysis of large amounts of data. Machine learning approaches play an important role in this development that has really just begun.

**Acknowledgements**

The author thanks BiGmax, the Max Planck Society's Research Network on Big-Data-Driven Materials-Science, for support and stimulating interdisciplinary interactions with members of the consortium.